\documentclass[aip,jcp,superscriptaddress,amsmath,amssymb,twocolumn,reprint]{revtex4-1}
\usepackage[version=3]{mhchem} 
\usepackage{graphicx}
\usepackage{amsmath}
\usepackage{amssymb}
\usepackage{nicefrac}
\usepackage{booktabs}

\usepackage{array}
\newcolumntype{m}{>{$} c <{$}}
\usepackage{color}

\def\rv{{\bf r}}
\def\fv{{\bf f}}

\def\beq{\begin{equation}}
\def\eeq{\end{equation}}

\begin{document}
    
\author{Stefan Vuckovic}
\affiliation
{Department of Theoretical Chemistry and Amsterdam Center for Multiscale Modeling, FEW, Vrije Universiteit, De Boelelaan 1083, 1081HV Amsterdam, The Netherlands}
\author{Tom Irons}
\affiliation
{School of Chemistry, University of Nottingham, University Park, Nottingham NG7 2RD, United Kingdom}
\author{Andreas Savin}
\affiliation
{Sorbonne Universit\'es, UPMC Univ Paris 06, UMR 7616, Laboratoire de Chimie Th\'eorique, F-75005 Paris, France} 
\affiliation
{CNRS, UMR 7616, Laboratoire de Chimie Th\'eorique, F-75005, Paris, France}
\author{Andrew M. Teale}
\affiliation
{School of Chemistry, University of Nottingham, University Park, Nottingham NG7 2RD, United Kingdom}
\author{Paola Gori-Giorgi}
\affiliation
{Department of Theoretical Chemistry and Amsterdam Center for Multiscale Modeling, FEW, Vrije Universiteit, De Boelelaan 1083, 1081HV Amsterdam, The Netherlands}
\email{p.gorigiorgi@vu.nl}

\title{Exchange--correlation functionals via local interpolation along the adiabatic connection}

\begin{abstract}
The construction of density-functional approximations is explored by modeling the adiabatic connection {\em locally}, using energy densities defined in terms of the electrostatic potential of the exchange-correlation hole. These local models are more amenable to the construction of size-consistent approximations than their global counterparts. In this work we use accurate input local ingredients to assess the accuracy of range of local interpolation models against accurate exchange-correlation energy densities. The importance of the strictly-correlated electrons (SCE) functional describing the strong coupling limit is emphasized, enabling the corresponding interpolated functionals to treat strong correlation effects. In addition to exploring the performance of such models numerically for the helium and beryllium isoelectronic series and the dissociation of the hydrogen molecule, an approximate analytic model is presented for the initial slope of the local adiabatic connection. Comparisons are made with approaches based on global models and prospects for future approximations based on the local adiabatic connection are discussed.
\end{abstract}
\maketitle

\section{Introduction}\label{sec_introedens}
Kohn--Sham density-functional theory (KS DFT)\cite{KohSha-PR-65} is the method most widely used in electronic structure calculations, due to its modest computational cost combined with an accuracy that is often competitive with much more expensive \textit{ab initio} methods. The accuracy of the method is limited by the quality of approximations required to describe the quantum mechanical exchange and correlation (XC) interactions of electrons. A large number of density functional approximations (DFAs) for the XC--energy have been developed in recent decades. 

The simplest DFAs are based on the local density approximation (LDA), as proposed by KS in their 1965 paper,\cite{KohSha-PR-65} in which the XC--energy is approximated as a functional of the density at a given point in space. The generalised gradient approximations (GGAs)\cite{LeeYanPar-PRB-88,Bec-PRA-88,PerCheVosJacPedSinFio-PRB-92,PerBurErn-PRL-96,ZhaTru-JCP-08} go beyond the LDA by modelling the XC--energy as a functional of the local density and its first derivative. The meta-GGAs\cite{TaoPerStaScu-PRL-03,PevTru-PTRSA-14,SunRuzPer-PRL-15} are closely related but their functional forms are also dependent on the KS kinetic energy density and/or, less commonly, the Laplacian of the electron density. Further developments led to the introduction of the occupied KS orbitals as ingredients for the XC energy (hybrid functionals,\cite{Bec-JCP-93,PerErnBur-JCP-96,ZhaSchTru-JCTC-06} self-interaction corrections,\cite{PerZun-PRB-81,VydScu-JCP-04,VydScu-JCP-05,KumKro-RMP-08}), and more recently also the virtual KS orbitals (double--hybrid functionals,\cite{Gri-JCP-06,LarGri-JCTC-10,ShaTouSav-JCP-11} random-phase approximations\cite{Fur-PRB-01,HessGor-MP-10,AngLiuTouJan-JCTC-11}). Local hybrid functionals\cite{JarScuErn-JCP-03,ArbKau-CPL-07,ArbKau-JCP-08,BahKau-JCTC-15} are also an interesting alternative approach to construct hybrid methods that are pertinent to the context of this work. 

The inclusion of additional dependencies in XC--functionals has often resulted in significant improvements in their accuracy for general calculations. However, these improvements cannot be described as systematic in the same way that the accuracy of an \textit{ab initio} calculation may be systematically improved by considering a larger number of excited determinants; some DFAs give excellent results for particular systems but perform very poorly otherwise, and \textit{vice versa}. There also remain many problems that none of the currently available DFAs can accurately model. An important example of this, which is pertinent to this work, are \textit{strong correlation} effects, commonly found in systems with near--degenerate orbitals such as the $d$-- and $f$--block elements, but also in systems where chemical bonds are being broken or formed. 

In the present work, the problem of constructing DFAs accurate for systems with and without strong correlation is examined by considering the adiabatic connection (AC)\cite{Har-PRA-84,LanPer-SSC-75} at the local level, i.e., in each point of space.\cite{MirSeiGor-JCTC-12} The AC, discussed in subsection~\ref{subsec_adcon}, 
provides an exact expression for the exchange and correlation energies by considering the changes that occur as the strength of electron interaction is smoothly increased from zero. This formalism has provided the basis for the development of several DFAs,\cite{Ern-CPL-96,BurErnPer-CPL-97,Bec-JCP-93,MorCohYan-JCPa-06} which attempt to interpolate the AC between the non--interacting and physical systems in order to estimate the XC--energy. An advantage of the AC formalism crucial to our construction, is that it allows the problem of strong correlation to be addressed in a more direct way, by creating interpolation models that are explicitly dependent on the strongly--interacting limit, in addition to the non--interacting limit, of the AC. 

The strongly--interacting limit of the AC has recently become the subject of much interest.\cite{SeiGorSav-PRA-07,GorVigSei-JCTC-09,MalGor-PRL-12,MirSeiGor-JCTC-12,MirSeiGor-PRL-13,VucWagMirGor-JCTC} The properties of the AC integrand in this limit reveal highly non--local density dependence of correlation effects\cite{Sei-PRA-99,SeiGorSav-PRA-07,MalGor-PRL-12,MalMirCreReiGor-PRB-13,MenMalGor-PRB-14} that cannot be obtained from the standard (semi)local or hybrid functionals. Study of the strongly--interacting limit in DFT has focused on \textit{strictly-correlated electrons} (SCE) functional, where the electrons have an infinite interaction strength. This limit is of particular interest from a theoretical point of view as it can be studied exactly for one--dimensional systems\cite{ColDepDiM-CJM-14} and may be closely approximated in systems with spherical or cylindrical symmetry.\cite{SeiGorSav-PRA-07,VucWagMirGor-JCTC-15} These studies show that in the limit of infinite interaction strength certain {\em integrals} of the density appear in the exchange-correlation functionals, revealing a mathematical structure very different from the one of the usual semi-local or orbital-dependent approximations. The nonlocal radius (NLR) functional proposed in ref \citenum{WagGor-PRA-14} approximates the SCE functional with a model that retains some of the SCE nonlocality, introducing the integrals of the spherically averaged density around a reference electron. The inclusion of the NLR functional into global and local interpolations along the adiabatic connection has been very recently explored by Zhou, Bahmann and Ernzerhof.\cite{ZhoBahErn-JCP-15} In another recent work, Kong and Proynov constructed a functional combining the information from the Becke'13 model\cite{Bec-JCP-13a} and approximating local quantities along the AC.\cite{KonPro-JCTC-15}

The aim of the present work is to start a systematic study of local interpolation models along the adiabatic connection, using at a first stage exact input ingredients, thus disentangling the errors due to the interpolation models from those due to the approximate ingredients. The local AC for several closed--shell atoms has been recently computed \cite{IroTea-MP-15} to high accuracy between the non--interacting and physical systems using the Legendre--Fenchel formulation of DFT due to Lieb,\cite{Lie-IJQC-83} and the Lieb maximisation method of refs\citenum{WuYan-JCP-03,TeaCorHel-JCP-09,TeaCorHel-JCP-10,TeaCorHel2-JCP-10}. Local information pertaining to the strongly--interacting limit is calculated using the SCE functional, and together these quantities are used to calculate local analogues of some established global AC interpolation functionals. We also discuss how to approximate crucial local ingredients such as the initial slope of the local adiabatic curve.

In section~\ref{sec_theo}, relevant theoretical background is given including an overview of the AC formalism and the construction of DFAs from both global and local variants of the AC. Techniques for computing quantities along the AC are discussed, including the determination of the local AC as introduced in ref~\citenum{IroTea-MP-15}. In section~\ref{loc_quant} the construction of a local model for the AC is discussed, considering the non-interacting and strong-interaction limits carefully in this context. The role of the SCE in constructing local AC interpolation models is examined. Finally the forms of some local interpolation models, taken from successful existing global models are introduced. In section~\ref{sec_results} the performance of these models is assessed for the helium and beryllium isoelectronic series and for dissociation of the H$_2$ molecule, a system that typifies the failure of present DFAs to properly account for strong--correlation. Directions for future work are outlined in section~\ref{sec_concl}. 

\section{Theoretical Background}\label{sec_theo}
\subsection{The Adiabatic Connection}\label{subsec_adcon}
The AC was proposed in a series of papers,\cite{HarJon-JPF-74,LanPer-SSC-75,GunLun-PRB-76,Har-PRA-84} which suggested that further insight into electronic correlations in DFT may be gained by considering a system at constant electron density as the interaction strength is smoothly scaled between zero, i.e. the KS auxiliary system, and the full physical interaction strength. This scaling of the interaction strength is achieved by the introduction of a simple \textit{coupling--constant} coefficient $\lambda$, such that the Hamiltonian for any given $\lambda$ is written as
\begin{equation}\label{eq:hamil_ac}
\hat{H}_{\lambda} = \hat{T} + \lambda\hat{W} + \hat{V}_\lambda,
\end{equation}
where $\hat{T}$ is the kinetic energy operator, $\hat{W}$ is the physical electron interaction operator and $\hat{V}_\lambda$ is the operator representing an external potential $v_{\lambda}$ that binds the electron density at $\lambda$, such that it is always equal to the density of the physically interacting system ($\rho_{\lambda} = \rho_1, \hspace{0.1in} \forall \lambda$). As the value of $\lambda$ is smoothly increased from zero to one, the system evolves adiabatically through a family of $\lambda$--dependent wave functions $\Psi_{\lambda}$ to the physical system described by $\Psi_{1}$. 

Given a Hamiltonian $\hat{H}_\lambda$, one can define the corresponding $\lambda$--dependant universal density functional as 
\begin{subequations}\begin{align}
F_{\lambda} [\rho] &= \min_{\Psi_{\lambda} \to \rho} \langle \Psi_{\lambda} |\hat{T} + \lambda\hat{W} | \Psi_{\lambda} \rangle \label{eq:funcrho_lam} \\[1ex]
&= F_{0} [\rho] + \int_{0}^{\lambda} \frac{\partial F_{\nu}}{\partial \nu}\, \mathrm{d}\nu, \label{eq:funcrho_lam_helfey} 
\end{align}\end{subequations}
where eq~\ref{eq:funcrho_lam_helfey} follows from the application of the Hellmann--Feynman theorem to eq~\ref{eq:funcrho_lam}. This allows the well--known AC formula to be derived, yielding the following exact expression for the XC--energy of an electronic system,
\begin{equation}\label{eq:ac_xc}
E_{\rm xc}[\rho] = \int_{0}^{1} \mathcal{W}_\lambda[\rho] \, \mathrm{d}\lambda,
\end{equation}
where $\mathcal{W}_{\lambda}[\rho]$ is the (global) AC integrand, given by
\begin{equation}\label{eq:w_lam}
\mathcal{W}_\lambda[\rho]= \langle \Psi_{\lambda}  |\hat{W}| \Psi_{\lambda} \rangle - U[\rho],
\end{equation}
$\Psi_{\lambda}[\rho]$ is the ground state wavefunction of $\hat{H}_{\lambda}$ in eq~\ref{eq:hamil_ac}, and $U[\rho]$ the Hartree (Coulomb) energy. 

The AC integrand may be characterized by several features that can be exactly defined: the expansion of $\mathcal{W}_{\lambda}$ in the non--interacting limit is given by\cite{GorLev-PRB-93}
\begin{equation}\label{eq:nilimit}
\mathcal{W}_\lambda[\rho] = \mathcal{W}_0 [\rho] + \mathcal{W}'_0[\rho] \lambda + \ldots \quad (\lambda \to 0),	
\end{equation}
whilst its expansion in the strongly--interacting limit can be expressed as\cite{Sei-PRA-99,SeiGorSav-PRA-07,GorVigSei-JCTC-09}
\begin{equation}\label{eq:w_ld} \begin{aligned}
\mathcal{W}_\lambda[\rho] &= \mathcal{W}_\infty[\rho] + \mathcal{W}'_\infty[\rho] \lambda^{-1/2} \\
&+ \mathcal{O}(\lambda^{-n}) \quad (\lambda \to \infty \,\, , \,\, n\ge 5/4).
\end{aligned}\end{equation}
Here, the non--interacting terms $\mathcal{W}_{0} [\rho]$ and $\mathcal{W}'_{0}[\rho]$ are the exchange energy and twice the second--order correlation energy given by G{\"o}rling-Levy perturbation theory (GL2)\cite{GorLev-PRB-93,GorLev-PRA-94} (see section~\ref{sec:ecGL2MP2}), respectively. Their analogues at the strongly--interacting limit, $\mathcal{W}_{\infty}[\rho]$ and $\mathcal{W}'_{\infty}[\rho]$, have been studied in Refs.~\citenum{Sei-PRA-99,SeiGorSav-PRA-07,GorVigSei-JCTC-09} and will be discussed further in section~\ref{sec_sce}. In addition to these asymptotic limits, the behaviour of $\mathcal{W}_{\lambda}$ under uniform coordinate scaling is also well--defined, as discussed in ref~\citenum{Lev-PRA-91}.

\subsection{DFAs based on the global adiabatic connection}
To construct practical DFAs one could consider modelling the integrand of eq~\ref{eq:w_lam} using a function that interpolates between the limits of eq~\ref{eq:nilimit} and eq~\ref{eq:w_ld}. The SCE limit is of particular importance in the present work, however, one could also consider models that intercept any other known point on the adiabatic connection for $\lambda > 1$. Several attempts to develop DFAs based on these ideas have been put forward in the literature, see e.g. Refs.~\citenum{Ern-CPL-96,BurErnPer-CPL-97,SeiPerLev-PRA-99,SeiPerKur-PRL-00,CohMorYan-JCP-07,SeiGorSav-PRA-07,GorVigSei-JCTC-09,LiuBur-PRA-09,Teale2010}. Each form makes a choice of a simple model function and the parameters on which to base the model. These parameters often include the known exact expressions for the parameters in eq~\ref{eq:nilimit}: $\mathcal{W}_{0}[\rho] = E_{\rm x}[\{ \phi_i \}]$, $\mathcal{W}^\prime_{0}[\rho] = 2 E_\text{c}^\text{GL2}[\{ \phi_p ,\epsilon_p\}]$, since these may be computed from the set of Kohn--Sham orbitals ($\{ \phi_p \}$) and orbital energies ($\{ \epsilon_p \}$). 

The calculation of $\mathcal{W}^{\prime}_{0}[\rho] = 2 E_\text{c}^\text{GL2}[\{ \phi_{p},\epsilon_{p}\}]$ requires the GL2 correlation energy,\cite{GorLev-PRB-93,GorLev-PRA-94} which leads to a computational cost similar to the second-order M{\o}ller--Plesset (MP2) model used in \textit{ab initio} quantum chemistry.\cite{Moller1934} The parameters in the SCE limit entering eq~\ref{eq:w_ld} are clearly also of special interest in this context, and they can be computed numerically for atomic systems and molecules with cylindrical symmetry.\cite{SeiGorSav-PRA-07,DiMGerNenSeiGor-xxx-15,
VucWagMirGor-JCTC-15} More frequently DFAs are derived for points along the AC with $\lambda>0$, often by employing scaling relations to derive forms from existing DFAs. A similar strategy can also be used to approximate $\mathcal{W}^\prime_0[\rho]$ by a DFA, see for example ref~\citenum{CohMorYan-JCP-07}.


In tandem with choosing a set of exact or approximate values to parameterize a model for the AC one must also choose an appropriate model function for the AC integrand. A number of these have been suggested and many have been benchmarked in practical applications. One of the simplest (and most often used) is that of a  [1/1] Pad{\'e}, as suggested by Ernzerhof.\cite{Ern-CPL-96} A range of forms were suggested by Cohen et al. and tested using approximate parameterizations,\cite{CohMorYan-JCP-07} leading to the MCY1 functional in which a [1/1] Pad{\'e} model is employed. Peach et al.\cite{PeaTeaToz-JCP-07,PeaMilTeaToz-JCP-08} attempted to disentangle approximations in the choice of parameters from those in the choice of model AC function by utilizing nearly exact KS orbitals and orbital energies derived from full configuration interaction data to calculate $\mathcal{W}_0[\rho]$ and $\mathcal{W}^\prime_0[\rho]$ and the corresponding interacting wave functions to evaluate $\mathcal{W}_1[\rho]$ via eq~\ref{eq:w_lam}. Our present study follows a similar philosophy, but applied to local, rather than global, interpolations.

Seidl and co workers\cite{SeiPerLev-PRA-99,SeiPerKur-PRL-00} were the first to make use of the strong-interaction limit (although approximated at a semilocal level, using the so-called point-charge-plus-continuum, or PC, functional) in constructing a global AC model, known as the interaction strength interpolation (ISI) functional. The revISI model\cite{GorVigSei-JCTC-09} and the models of Liu and Burke\cite{LiuBur-PRA-09} were later constructed to take account of the $\lambda^{-1/2}$ dependence of the second term of eq~\ref{eq:w_ld}, which was not correctly described by the ISI approach. Teale, Coriani and Helgaker also proposed forms for the AC integrand based on the structure of traditional \textit{ab initio} methodologies\cite{Teale2010} and parameterized these forms to intercept values of the AC at any $\lambda>0$. 

The majority of these models for the global AC suffer from the fact they are not size consistent in practice. This deficiency arises from a non-linear dependence on the parameters $\mathcal{W}_0$, $\mathcal{W}_0^\prime$, and a chosen approximation to $\mathcal{W}_{(\lambda>0)}$. When these global parameters enter in a non-linear fashion (often as ratios) then size consistency is difficult to achieve. One route forward is to construct local AC models, which can replace these global parameters with local values defined at each point in space and may be more amenable to the construction of models that recover size-consistency (at least in the usual density-functional sense\cite{GorSav-JPCS-08,Sav-CP-09}).   

\subsection{Constructing a local adiabatic connection}\label{loc_quant}
The AC expression for the XC--energy of a system as given in eq~\ref{eq:ac_xc} describes a \textit{global} quantity, integrated over the coupling--constant $\lambda$. However, it may equally be written as the spatial integral of a \textit{local} quantity analogous to the local value of an XC--functional. To this effect, eq~\ref{eq:ac_xc} may be re--written as
\begin{equation}\label{eq:ac_wxc}
E_{\rm xc}[\rho] = \int_{0}^{1} {\rm d}\lambda\int {\rm d}\mathbf{r} \rho(\mathbf{r}) w_\lambda(\mathbf{r}),
\end{equation}
where $w_\lambda(\mathbf{r})$ is the energy density at a given $\lambda$. It is well known that the energy density cannot be uniquely defined;\cite{BurCruLam-JCP-88,CruLamBur-JPCA-98,TaoStaScuPer-PRA-08} an arbitrary number of terms may be added to $w_\lambda(\mathbf{r})$, yet an identical $\mathcal{W}_\lambda[\rho]$ recovered if their spatial integral is zero. Thus any such energy densities are only defined within a particular \textit{gauge}, and only energy densities defined in the same gauge may be meaningfully compared. 

In the context of the present work, it is both convenient and physically meaningful to define $w_{\rm xc, \lambda}(\mathbf{r})$ in the gauge of the electrostatic potential of the exchange--correlation hole,
\begin{equation}\label{eq:wxc_def}
w_\lambda(\mathbf{r}) = \frac{1}{2} \int \frac{h_{\rm xc}^{\lambda}(\mathbf{r},\mathbf{r}')}{|\mathbf{r} - \mathbf{r}'|} \, \mathrm{d} \mathbf{r}'
\end{equation}
where $h_{\rm xc}^{\lambda}(\mathbf{r},\mathbf{r}')$ is the exchange--correlation hole,
\begin{equation}\label{eq:hxc_def}
h_{\rm xc}^{\lambda}(\mathbf{r},\mathbf{r}') = \frac{P_{2}^{\lambda} (\mathbf{r},\mathbf{r}')}{\rho(\mathbf{r})} - \rho(\mathbf{r}'),
\end{equation}
and $P_{2}^{\lambda}(\mathbf{r},\mathbf{r}')$ is the pair--density obtained from the wave function $\Psi_{\lambda}[\rho]$
\begin{equation}\label{eq:pair}\begin{aligned}
P_{2}^{\lambda}&(\mathbf{r},\mathbf{r}') = N(N-1) \times \\
& \sum_{\sigma_{1}\ldots\sigma_{N}} \int |\Psi_{\lambda}(\mathbf{r}\sigma_{1},\ldots,\mathbf{r}_{N}\sigma_{N})|^{2} \, \mathrm{d} \mathbf{r}_{3} \ldots \mathrm{d} \mathbf{r}_{N}.
\end{aligned}\end{equation}
The definition of energy densities in the gauge of the XC--hole is well--established in the literature, and further discussion may be found in refs\citenum{BurCruLam-JCP-98,Armiento2002,MirSeiGor-JCTC-12}. 
The coupling--constant averaged ($\lambda$--averaged) XC--energy density is defined as
\begin{equation}\label{eq:wxc_cca}
\bar{w}_{\rm xc}(\mathbf{r}) = \int_{0}^{1} w_\lambda(\mathbf{r}) \mathrm{d} \lambda.
\end{equation} 
As the spatial integral of the product of this quantity and the density yields the XC--energy, the same quantity may be considered as a target to be modelled by XC--functionals,\cite{IroTea-MP-15} although GGAs and metaGGAs often aim at energy densities within different definitions.\cite{BurCruLam-JCP-98,PerBurErn-PRL-96,TaoPerStaScu-PRL-03} Given the invariance of the exchange energy to electron--interaction strength, eq~\ref{eq:wxc_cca} may be trivially resolved into separate exchange and correlation terms as
\begin{equation}\label{eq:wxc_wxwc}\begin{aligned}
\bar{w}_{\rm c}(\mathbf{r}) &= \bar{w}_{\rm xc}(\mathbf{r}) - \bar{w}_{\rm x}(\mathbf{r}) \\
&= \bar{w}_{\rm xc}(\mathbf{r}) - w_{\lambda=0}(\mathbf{r})
\end{aligned}\end{equation}
The aim of the local interpolation schemes examined in this work is to approximate $\bar{w}_{\rm xc}(\mathbf{r})$ and $\bar{w}_{\rm c}(\mathbf{r})$ through interpolating the local AC. In principle, this approach is analogous to that of the global AC interpolation schemes previously discussed, but rather than depending on quantities pertaining to the global AC, they are instead constructed from their local equivalents, $w_\lambda(\mathbf{r})$. Obviously, a local interpolation
will only be meaningful if all of the local terms are defined in the same gauge. It is again both convenient and logical to define all local quantities in the gauge of eq~\ref{eq:wxc_def}, as in which highly accurate energy densities $w_\lambda(\mathbf{r})$ in the range $0 \leq \lambda \leq 1$ have previously been calculated,\cite{IroTea-MP-15} and additionally can be computed for small systems in the limit $\lambda \to \infty$.\cite{MirSeiGor-JCTC-12,VucWagMirGor-JCTC} 

At $\lambda=0$, the energy density in the gauge of eq~\ref{eq:wxc_def} is the exchange energy density $w_0(\mathbf{r})=w_x(\mathbf{r})$, often denoted $\epsilon_x(\mathbf{r})$ in the literature (also equal to $1/2$ the non-local Slater potential\cite{Sla-PR-51}), which is the crucial ingredient of local hybrid functionals. Accurate and efficient computational schemes for this quantity have become available in the recent years.\cite{BahKau-JCTC-15,JanKruScu-JCP-08} In a way, local interpolation models can be viewed as local hybrids that carefully address the gauge problem. 

The local equivalent of $\mathcal{W}'_{0}$ is not as simple to define, yet is an essential component of AC interpolation schemes as it provides a measure of the departure from exchange-only behaviour, in other words provides the information from which the correlation energy is approximated. Whilst many global models use GL2 theory for this purpose, its dependence on global quantities makes it unclear how it could be applied to a local interpolation scheme. This is discussed in detail in section~\ref{sec:ecGL2MP2}.  

\subsection{The Lieb Maximization}\label{liebmax}
In order to assess the quality of our local interpolation functionals, it is necessary to have accurate data of energy--densities, defined in the gauge of the XC--hole. These may be acquired by the method of Lieb maximisation, described in Refs.~\citenum{TeaCorHel-JCP-09,TeaCorHel-JCP-10,TeaCorHel2-JCP-10}. 

The Lieb maximisation is an optimisation algorithm developed using the convex--conjugate functional defined by Lieb in ref~\citenum{Lie-IJQC-83} as the Legendre--Fenchel transform to the energy,
\begin{subequations}\begin{align}
E_{\lambda}[v] &= \inf_{\rho} \, \, \left\lbrace F_{\lambda}[\rho] + \int v(\mathbf{r}) \rho(\mathbf{r}) \, {\rm d}\mathbf{r} \right\rbrace \label{eq:lieb_ev} \\[1ex]
F_{\lambda}[\rho] &= \sup_{v} \left\lbrace E_{\lambda}[v] - \int v(\mathbf{r}) \rho(\mathbf{r}) \, {\rm d}\mathbf{r} \right\rbrace \label{eq:lieb_fp} 
\end{align}\end{subequations}
in which the density $\rho$ and potential $v$ are conjugate variables, belonging to the dual vector spaces
\begin{equation}\label{eq:dual_space}
\rho \in L^{3} \cap L^{1} \qquad \qquad v \in L^{\frac{3}{2}} + L^{\infty}
\end{equation}
and $E_{\lambda}[v]$ is the energy yielded by a given electronic structure calculation at potential $v(\mathbf{r})$. This convex--conjugate formulation follows from the concavity of variationally-determined energy $E_{\lambda}[v]$ in $v$, from which Lieb showed that $F_{\lambda}[\rho]$ must be convex in $\rho$. Furthermore the conjugate functional to a nonconcave energy, such as that which may result from a non--variational calculation, remains well--defined as it is necessarily convex. A subsequent Legendre--Fenchel transform of $F_{\lambda}[\rho]$ yields the concave envelope (least concave upper bound) to $E_{\lambda}[v]$, hence unique solutions to $F_{\lambda}[\rho]$ can always be obtained.

In the Lieb maximisation, the optimised density $\rho(\mathbf{r})$ is obtained by maximising $F_{\lambda}[\rho]$ with respect to variations in the potential $v(\mathbf{r})$, rather than by minimising $E_{\lambda}[v]$ with respect to $\rho(\mathbf{r})$ as is usually the case. Therefore at convergence, the potential $v(\mathbf{r})$ in eq~\ref{eq:lieb_fp} is that which yields $\rho(\mathbf{r})$. In the present work, Lieb maximisations have been carried--out at a number of points along the AC in the range $0 \leq \lambda \leq 1$, hence the density is constrained such that $\rho_{\lambda} = \rho_{\lambda=1}$, resulting in a $\lambda$--dependent optimizing potential. 

In order to effectively optimize with respect to the potential, we parameterize it by using the method of Wu and Yang (WY) \cite{WuYan-JCP-03} as
\begin{equation}\label{eq:wuyang}
v(\mathbf{r}) = v_{\rm ext}(\mathbf{r}) + (1 - \lambda) v_{\rm ref}(\mathbf{r}) + \sum_{t} b_{t} g_{t}(\mathbf{r}),
\end{equation}
where $v_{\rm ext}(\mathbf{r})$ is the external potential due to nuclei, $v_{\rm ref}(\mathbf{r})$ is a reference potential chosen to ensure that $v(\mathbf{r})$ has the correct asymptotic behaviour, and $\left\lbrace g_{t} \right\rbrace$ are a set of Gaussian functions with coefficients $\left\lbrace b_{t} \right\rbrace$. In all calculations in this work we choose the potential expansion basis set to be identical to the primary orbital basis set. The reference potential used in this work is the Fermi--Amaldi potential\cite{Fermi1934} and $F_{\lambda}[\rho]$ is optimized with respect to the coefficients of the potential basis $\left\lbrace b_{t} \right\rbrace$. Additionally, convergence is accelerated through the use of the Newton method described in Refs.~\citenum{TeaCorHel-JCP-09,TeaCorHel-JCP-10,TeaCorHel2-JCP-10}. The relaxed--Lagrangian formulation of Helgaker and J{\o}rgensen\cite{Helgaker1989} is used to obtain relaxed densities for non--variational wavefunctions, which serve as input to the Lieb functional and are used in the determination of the derivatives required for its optimization. 

In this work, Lieb maximisation calculations are performed using the implementation of Refs.~\citenum{TeaCorHel-JCP-09,TeaCorHel-JCP-10,TeaCorHel2-JCP-10} in a development version of the \textsc{Dalton} quantum chemistry software package,\cite{Dalton-WIRES-14} in which $E_{\lambda}[v]$ is computed by using coupled--cluster singles and doubles (CCSD)\cite{Purvis1982} and full configuration--interaction (FCI) wavefunctions. At convergence, where the optimising potential is such that $\rho_{\lambda} = \rho_{\lambda=1}$, the relaxed $\lambda$--interacting one-- and two--particle density matrices are computed, with which the $\lambda$-dependent XC energy densities may be obtained as
\begin{equation}\label{eq:lieb_wxc}
w_\lambda(\mathbf{r}) = \frac{1}{2\rho(\mathbf{r})} \int \frac{P_{2}^{\lambda}(\mathbf{r},\mathbf{r}')}{|\mathbf{r} - \mathbf{r}'|} {\rm d}\mathbf{r}' - \frac{1}{2} \int \frac{\rho(\mathbf{r}')}{|\mathbf{r} - \mathbf{r}'|} {\rm d}\mathbf{r}'.
\end{equation}

\section{Modelling the Local AC}
\subsection{The local slope in the non-interacting limit}\label{locslope}
As described in section~\ref{loc_quant}, the initial slope of the AC is an important part of many global AC models, in which it may be calculated directly by GL2 perturbation theory, however there is no analogous expression that yields the local equivalent and we give such an expression in section~\ref{sec:ecGL2MP2}. Here, the local initial slope of the XC energy density that is given in eq~\ref{eq:wxc_def} is defined as
\begin{equation}\label{eq:w'0}
w'_{0}(\mathbf{r}) = \left. \frac{\partial w_\lambda(\mathbf{r})}{\partial \lambda}\right|_{\lambda=0} \equiv \left. \frac{\partial w_{\rm c,\lambda}(\mathbf{r})}{\partial \lambda}\right|_{\lambda=0},
\end{equation}
and is related to the global slope, hence the GL2 correlation energy, by
\begin{equation}\label{eq:gl2w}
\mathcal{W}'_{0}[\rho]=\int w'_{0}(\mathbf{r}) \rho(\mathbf{r}) \, {\rm d}\mathbf{r}.
\end{equation} 

\subsubsection{Numerical calculation of the local slope}\label{num_slope}
In this study $w'_{0}(\mathbf{r})$ is numerically approximated by the method of finite difference, with a series of $w_\lambda(\mathbf{r})$ for $\lambda << 1$.

\begin{figure}
\includegraphics[width=\linewidth]{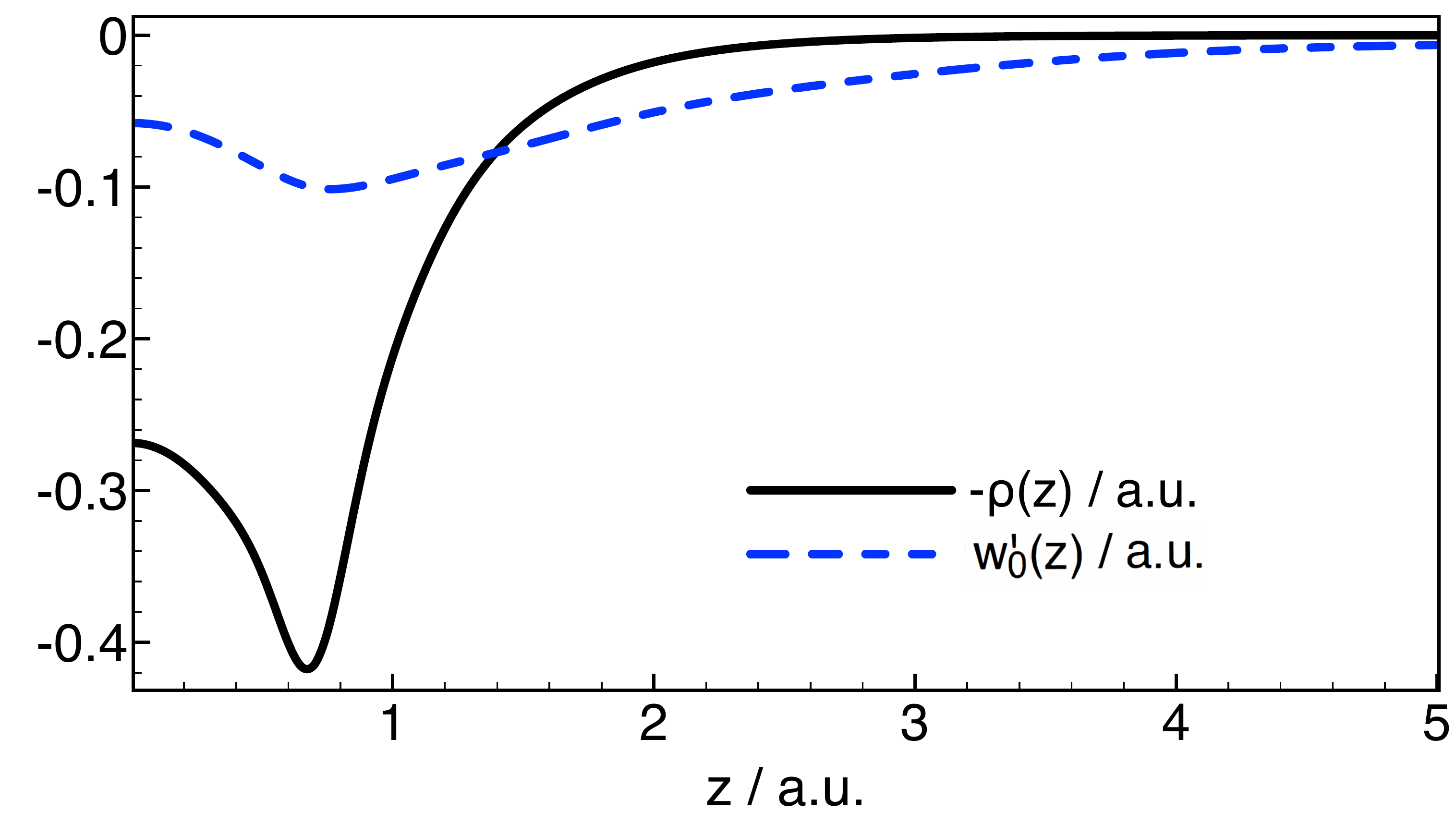}
\includegraphics[width=\linewidth]{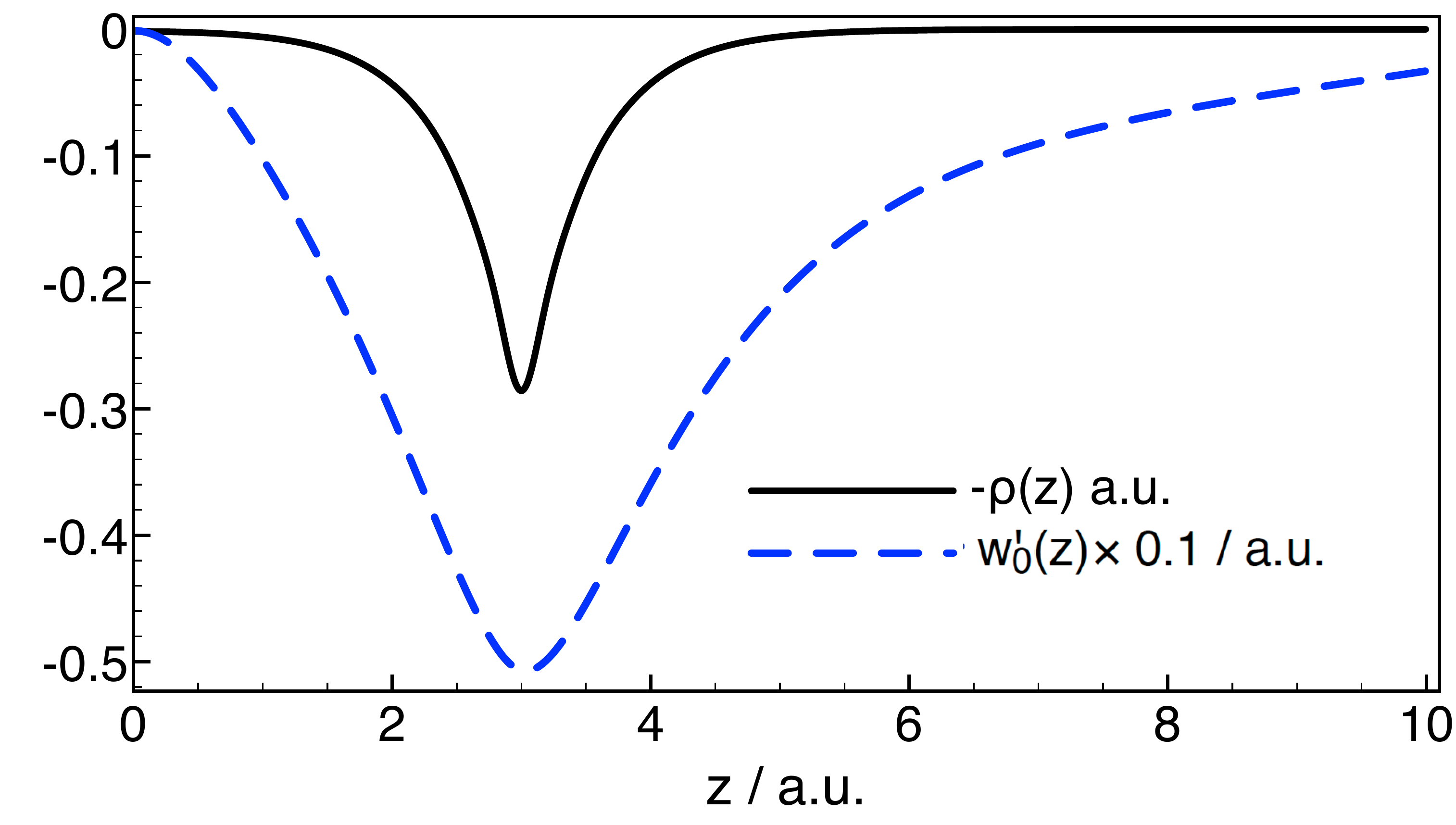}
\caption{Plots comparing the values of $-\rho(\mathbf{r})$ and $w'_{0}(\mathbf{r})$, with respect to the distance from the bond midpoint, $z$ / a.u., along the principal axis of the $\rm H_{2}$ molecule with bond lengths of $1.4$ a.u. (upper panel) and $6.0$ a.u. (lower panel).}
\label{fig_slopes}
\end{figure}

The resulting local slopes in the $\rm H_{2}$ molecule with bond length of $1.4$ a.u. and $6.0$ a.u. are plotted along the H--H bond in Figure~\ref{fig_slopes}, along with the densities from which they are calculated, at the FCI level of theory and in the uncontracted aug-cc-pCVTZ basis set.\cite{Dun-JCP-89} In both cases, the local slope is greatest in magnitude at the nuclei, as has been seen previously in atoms.\cite{IroTea-MP-15} It can be seen that the magnitude of the local slope is significantly larger in the stretched $\rm H_{2}$ molecule, mirroring observations previously made of the global AC in the dissociating hydrogen molecule.\cite{TeaCorHel-JCP-10}
\begin{figure}
\includegraphics[width=\linewidth]{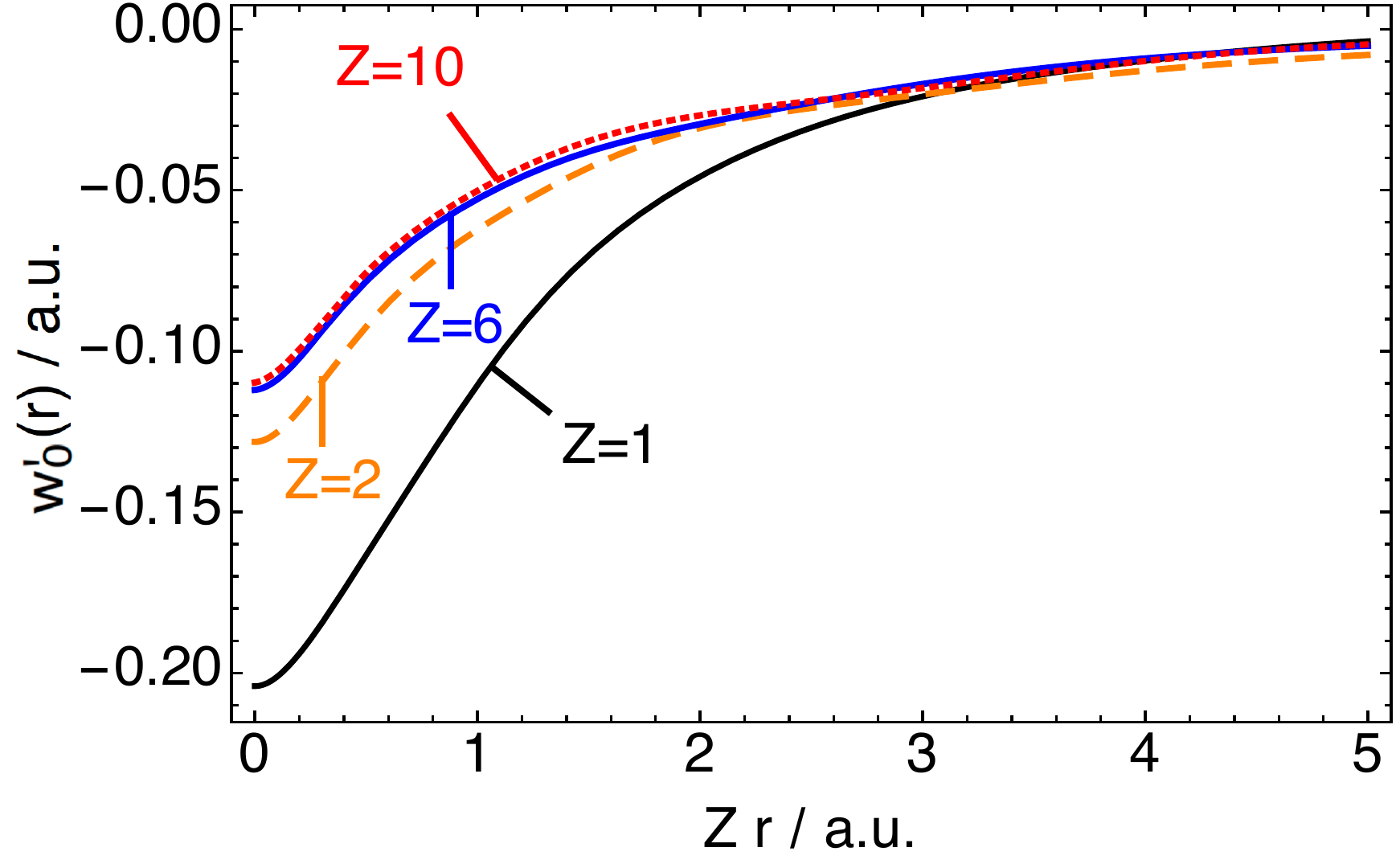}
\caption{Plots of $w'_{0}(\mathbf{r})$ for the helium isoelectronic series, with nuclear charges $1 \leq Z \leq 10$, and with radial distance from the nucleus $r$ / a.u. scaled by nuclear charge.}
\label{fig_slope_he}
\end{figure}

\begin{figure}
\includegraphics[width=\linewidth]{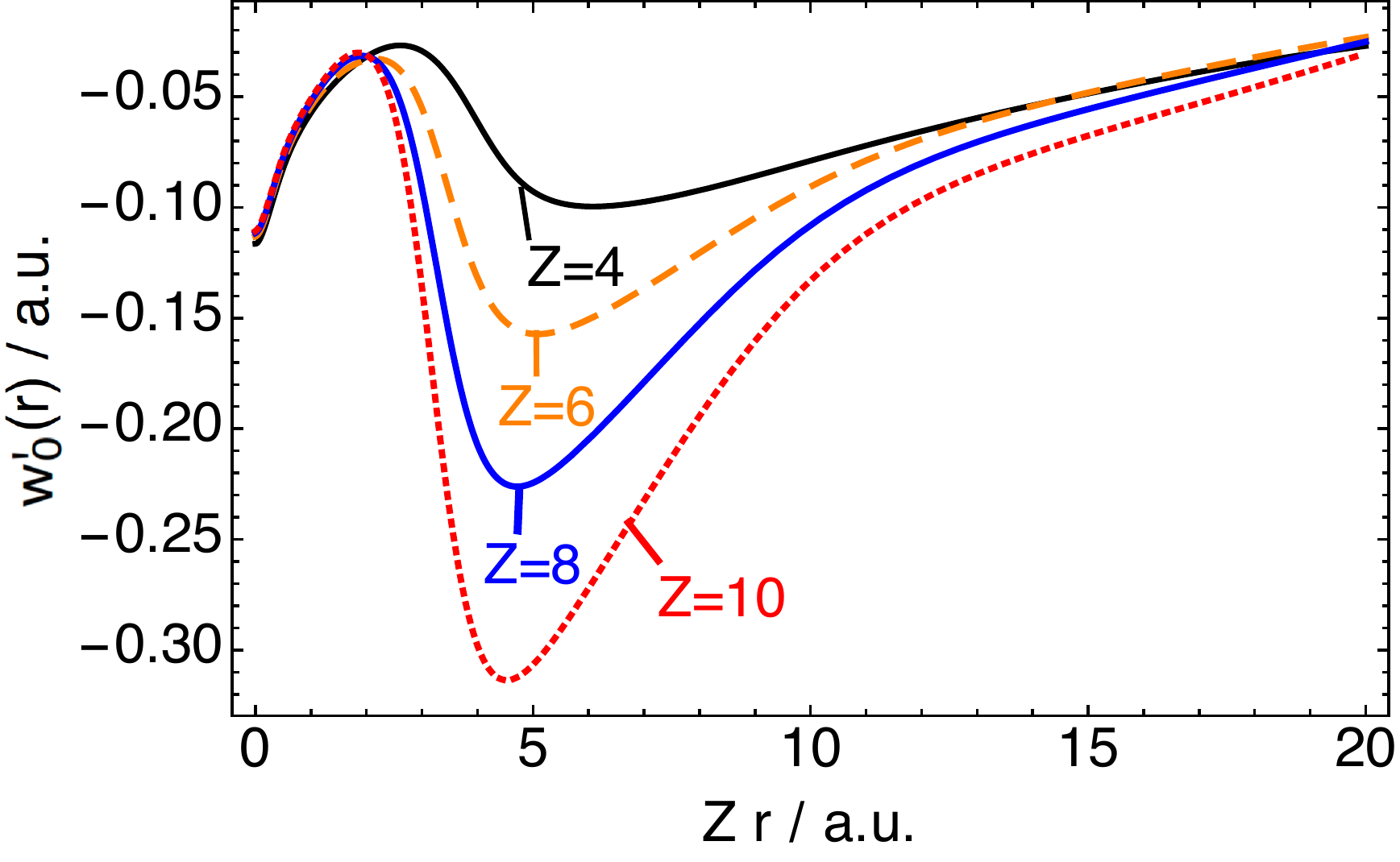}
\caption{Plots of $w'_{0}(\mathbf{r})$ for the beryllium isoelectronic series, with nuclear charges $4 \leq Z \leq 10$, and with radial distance from the nucleus $r$ / a.u. scaled by nuclear charge.}
\label{fig_slope_be}
\end{figure}

The local slopes in the He and Be isoelectronic series are plotted in Figures~\ref{fig_slope_he} and ~\ref{fig_slope_be} respectively. It is clear that, with increasing nuclear charge, the charge densities in both series become increaslingly contracted. The $x$--axis in both plots has been scaled by nuclear charge, highlighting a contrast in their behaviour with respect to the uniform scaling condition, 
\begin{equation}\label{eq:scal_rels}
\lim_{\gamma\to\infty} \frac{E_{\rm xc}[\rho_{\gamma}]}{E_{\rm x}[\rho_{\gamma}]}=1, \quad {\rm with}\;\;\rho_{\gamma}(\mathbf{r}) = \gamma^{3}\rho(\gamma\,\mathbf{r})
\end{equation}
which holds for non-degenerate KS systems.\cite{PerStaTaoScu-PRA-08} In Figure~\ref{fig_slope_he}, it can be seen that the slope of the AC for the He series becomes less negative with increasing $Z$, tending to an asymptotic value as $Z \to \infty$, consistent with the scaling relation of eq~\ref{eq:scal_rels}. However, the slope of the AC in the Be isoelectronic series becomes \textit{more} negative with increasing $Z$, indicating that the scaling relation is not satisfied by this series.\cite{ColSav-JCP-99}

\subsubsection{A functional approximation for the local slope}\label{sec:ecGL2MP2}
Whilst it is useful to numerically approximate the local slope for the purposes of evaluating local interpolation schemes, such functionals would only be viable for mainstream use in DFT calculations if they can be described by simple functional forms.

In global models, the initial slope can be calculated directly from the occupied and virtual KS orbitals according to GL2 theory,
\begin{equation}\label{eq:gl2}\begin{aligned}
\mathcal{W}_0^\prime[\rho] &= 2E_{c}^{\rm GL2}[\lbrace \phi_{p}, \epsilon_{p} \rbrace] \\
&= -\frac{1}{2}\sum_{abij}\frac{|\langle\phi_i \phi_j||\phi_a \phi_b \rangle|^{2}}{\epsilon_{a}+\epsilon_{b}-\epsilon_{i}-\epsilon_{j}} \\
&\hspace{0.16in}-2 \sum_{ia}\frac{|\langle \phi_{i}|\hat{v}_{\rm x}^{\rm KS}-\hat{v}_{\text{x}}^{\rm HF}|\phi_{a}\rangle|^{2}}{\epsilon_{a}-\epsilon_{i}},
\end{aligned}\end{equation}
where the indices $i,j$ and $a,b$ pertain to occupied and virtual KS orbitals respectively, $\hat{v}_{\rm x}^{\rm KS}$ is the local KS potential and $\hat{v}_\text{x}^{\rm HF}$ the non--local Hartree--Fock (HF) exchange potential. The first term in eq~\ref{eq:gl2} is analogous to the correlation energy given by MP2 theory, in which $\phi_p$ and $\epsilon_p$ are canonical HF orbitals and eigenvalues rather than KS ones. The second term accounts for the difference between the KS and HF exchange potentials and has a form similar to a singles term in many--body perturbation theory. Previous studies of GL2 theory have found that the second term, although non--negligible, is small in magnitude relative to the MP2--like term evaluated with the KS orbitals. \cite{TouShaBreAda-JCP-11} Therefore, it follows that an approximation to the GL2 correlation energy may be obtained by evaluating the MP2 correlation energy\cite{Gri-JCP-06} with the KS orbitals and eigenvalues, $E_{c}^{\rm GL2}\approx E_{c}^{\rm MP2}$.
 
Given that an approximation to the global AC slope may be obtained from an MP2--like calculation, it follows that an approximation to the local AC slope may be obtained by deriving a local form of this expression. Whilst MP2 theory treats perturbations of the wavefunction, the analysis may be extended to energy densities in the gauge of the XC--hole by means of eq~\ref{eq:pair}, as the substitution of eq~\ref{eq:lieb_wxc} into eq~\ref{eq:w'0} yields the following,
\begin{equation}\label{eq:wp2}
w'_{0}(\mathbf{r}) = \frac{1}{2\rho(\mathbf{r})} \int \frac{P'_{2}(\mathbf{r},\mathbf{r}')}{|\mathbf{r}-\mathbf{r}'|}\,{\rm d}\mathbf{r}',
\end{equation}
where $P'_{2}(\mathbf{r},\mathbf{r}')$ is the derivative of the pair--density at $\lambda=0$,
\begin{equation}\label{eq:p'2}
P'_{2}(\mathbf{r},\mathbf{r}') = \left. \frac{\partial P_{2,\lambda}(\mathbf{r},\mathbf{r}')}{\partial \lambda}\right|_{\lambda=0}.
\end{equation}
 Notice that eq~\ref{eq:wp2} ensures that $w'_0(\rv)$ is in the gauge of the electrostatic potential of the xc hole. Given a non--interacting ground--state wavefunction $\Psi^{(0)}$, the perturbed wavefunction $\Psi_{\lambda}$ for $|\lambda| << |\Psi^{(1)}-\Psi^{(0)}|^{2}$ can be appproximated by the series expansion
\begin{equation}\label{eq:ptpsi}
\Psi_{\lambda} = \sum_{n=0} \lambda^{n} \, \Psi^{(n)}.
\end{equation}
If one assumes that $\Psi^{(0)}$ is non--degenerate and has the form of a single Slater determinant, the first--order correction to the wavefunction is given by
\begin{equation}\label{eq:psi1}
\Psi^{(1)} = \sum_{k} \frac{\langle \Psi^{(0)}_{k} | \hat{W} - \hat{V}_{\rm HF} | \Psi^{(0)} \rangle}{E^{(0)}_{0}-E^{(0)}_{k}}\Psi^{(0)}_{k}.
\end{equation}
Restricting the space of $\Psi^{(0)}_{k}$ to doubly--excited determinants reduces this expression to
\begin{equation}\label{eq:psi1orb}
\Psi^{(1)} = -\sum_{abij} \frac{\langle \Psi_{ij}^{ab} | \hat{W} - \hat{V}_{\rm HF} | \Psi^{(0)} \rangle} {\epsilon_{a}+\epsilon_{b}-\epsilon_{i}-\epsilon_{j}}\Psi_{ij}^{ab}.
\end{equation}
In MP2 theory, contributions to the correlation energy from singly--excited determinants are necessarily zero due to Brillouin's theorem. However, this is not strictly true in GL2 theory as the singles term in eq~\ref{eq:gl2} makes a small, but non--zero, contribution to the GL2 correlation energy.\cite{GorLev-PRB-93} As such, considering only double--excitations in the model for the local slope can only yield approximations to the local slope; spatial integration of this quantity will not return the exact GL2 correlation energy. 

Application of the Slater-Condon rules to eq~\ref{eq:psi1orb} allows it to be re--written as
\begin{equation}\label{eq:psi1orbs}
\Psi^{(1)} = -\frac{1}{4}\sum_{abij} \frac{\langle ij||ab \rangle} {\epsilon_{a}+\epsilon_{b}-\epsilon_{i}-\epsilon_{j}}\Psi_{ij}^{ab},
\end{equation}
where the coefficient to $\Psi_{ij}^{ab}$ may be identified as an MP2 doubles--amplitude $t_{ij}^{ab}$, 
\begin{equation}\label{eq:mp2_tijab}
t_{ij}^{ab} = \frac{\langle ij||ab \rangle} {\epsilon_a+\epsilon_b-\epsilon_i-\epsilon_j}.
\end{equation}
To obtain $P'_{2}(\mathbf{r},\mathbf{r}')$, it is necessary to take the derivative of the pair--density corresponding to the perturbed wavefunction, $\Psi_{\lambda} \approx \Psi^{(0)} + \lambda \Psi^{(1)}$. Substituting this into eq~\ref{eq:pair} and rearranging the resulting expressions yields the following,
\begin{equation}\label{eq:pair_op}\begin{aligned}
P'_{2}(\mathbf{r},\mathbf{r}') &= N(N-1)\sum_{\sigma} \int \left. \frac{\partial |\Psi_{\lambda}|^{2}}{\partial \lambda}\right|_{\lambda=0} \, {\rm d}\mathbf{r}_{3} .. {\rm d}\mathbf{r}_{\rm N} \\
&= N(N-1)\sum_{\sigma} \int |2\Psi^{(0)} \Psi^{(1)}| \, {\rm d}\mathbf{r}_{3} .. {\rm d}\mathbf{r}_{\rm N} \\
&= 2 \left\langle \Psi^{(0)} \left| \hat{P}_{2}(\mathbf{r},\mathbf{r}') \right| \Psi^{(1)} \right\rangle ,
\end{aligned}\end{equation}
where we assume that $\Psi^{(0)}$ and  $\Psi^{(1)}$ are real and  $\hat{P}_{2}(\mathbf{r},\mathbf{r}')=N(N-1)\sum\limits_{i\neq j}\delta(\rv-\rv_i)\delta(\rv-\rv_j)$ is the pair--density operator. Substituting eq~\ref{eq:psi1orbs} into this expression gives,
\begin{equation}\label{eq:p'2w1}
P'_{2}(\mathbf{r},\mathbf{r}') = -\sum_{abij} t_{ij}^{ab} \langle \Psi^{(0)} |\hat{P}_{2}(\mathbf{r},\mathbf{r}')| \Psi_{ij}^{ab} \rangle,
\end{equation}
which may then be resolved into the following orbital--explicit expression,
\begin{equation}\label{eq:p'2orb}\begin{aligned}
P'_{2}(\mathbf{r},\mathbf{r}') = -2\sum_{abij} t_{ij}^{ab} &\left\lbrace \phi_i(\mathbf{r}) \phi_j(\mathbf{r}') \phi_a(\mathbf{r}) \phi_b(\mathbf{r}')\delta_{\sigma_i \sigma_a} \delta_{\sigma_j \sigma_b} \right.\\
- &\left. \phi_i(\mathbf{r}) \phi_j(\mathbf{r}') \phi_b(\mathbf{r}) \phi_a(\mathbf{r}')\delta_{\sigma_i \sigma_b} \delta_{\sigma_j \sigma_a} \right\rbrace,
\end{aligned}\end{equation}
where $\delta_{\sigma_i \sigma_a}$ is the Kronecker delta over two spin functions: $\int \sigma_i^*(m_s)\sigma_a(m_s) \mathrm{d}m_s=\delta_{\sigma_i \sigma_a}$. Substituting eq~\ref{eq:p'2orb} into eq~\ref{eq:wp2} finally results in an expression for the local slope,
\begin{equation}\label{eq:simplemp2}
w'_{0}(\mathbf{r}) = -\frac{1}{2\rho(\mathbf{r})} \sum_{abij} t_{ij}^{ab} v_{abij}(\mathbf{r}),
\end{equation}
where $v_{abij}(\mathbf{r})$ is the \textit{antisymmetrized orbital potential},
\begin{equation}\label{eq:orbpot}\begin{aligned}
v_{abij}(\mathbf{r}) &= \phi_i(\mathbf{r}) \phi_a(\mathbf{r}) \int \frac{\phi_j(\mathbf{r}') \phi_b(\mathbf{r}')}{|\mathbf{r} - \mathbf{r}'|}   \,\mathrm{d}\mathbf{r}' \delta_{\sigma_i \sigma_a} \delta_{\sigma_j \sigma_b}\\
&- \phi_i(\mathbf{r}) \phi_b(\mathbf{r}) \int \frac{\phi_j(\mathbf{r}')\phi_a(\mathbf{r}')}{|\mathbf{r}-\mathbf{r}'|}\,\mathrm{d}\mathbf{r}' \delta_{\sigma_i \sigma_b} \delta_{\sigma_j \sigma_a}.
\end{aligned}\end{equation}
Multiplying the right--hand side of eq~\ref{eq:simplemp2} by the density and integrating over all space, we recover twice the MP2 like expression. This is not an exact expression for the local slope, as the second term of eq~\ref{eq:gl2} is not accounted for. However, the omitted term is generally small relative to the MP2--like term, and vanishes entirely for two--electron systems, hence the expression for the local slope in eq~\ref{eq:simplemp2} should, in principle, be a fair approximation of the exact local slope. 

In future work we will implement and test eq~\ref{eq:simplemp2} against the numerical results in section~\ref{num_slope}. The doubles amplitudes $t_{ij}^{ab}$ are readily obtainable from standard quantum chemical packages, the potential $v_{abij}(\mathbf{r})$ can also be readily calculated; however, it would likely be computationally expensive to evaluate on a numerical grid. To reduce this cost a range of techniques, commonly used to accelerate the calculation of integrals in linear--scaling software packages, may be employed.\cite{Werner2003,Reine2008,DominguezSoria2009,BahKau-JCTC-15}
 
We note that the behaviour of the local slopes presented in Figures~\ref{fig_slopes},~\ref{fig_slope_he} and~\ref{fig_slope_be} may be rationalized in a similar manner to that commonly discussed for global models in terms of eq~\ref{eq:gl2}. This is because of the key role of the doubles amplitude $t_{ij}^{ab}$ in eq~\ref{eq:simplemp2}. The doubles amplitude has a dependence on the orbitals and orbital energies that is similar to that of the GL2 energy in eq~\ref{eq:gl2}. We see in Figure~\ref{fig_slopes} that the local slope of the hydrogen molecule displays the minima at the nuclei. Equation~\ref{eq:simplemp2}, which is exact for two--electron systems, can be used to rationalize this observation. For closed shell two--electron systems with only one virtual orbital, eq~\ref{eq:simplemp2} is simplified as follows:

\begin{equation}\label{eq:2elec2}\begin{aligned}
w'_{0}(\mathbf{r}) = -\frac{\langle 11 |22 \rangle \phi_1 (\mathbf{r})\phi_2 (\mathbf{r})}{\rho(\mathbf{r})(\epsilon_2-\epsilon_1)}\int\frac{\phi_1 (\mathbf{r}')\phi_2(\mathbf{r}')}{|\mathbf{r} - \mathbf{r}'|}d\mathbf{r}'.
\end{aligned}\end{equation}
Even if we used a minimal orbital basis for the evaluation of the expression given in eq~\ref{eq:2elec2} for the hydrogen molecule, we would see that the local slope is most negative at the two nuclei, for any bond length. Whilst this effect is captured with the minimal basis, the same minimal basis model would incorrectly describe the slope at at the bond midpoint. For example, in the top panel of Figure~\ref{fig_slopes} we see that $w'_0(\rv)$ is less than $0$ at the bond midpoint of H$_2$ at $R=1.4$. Within the minimal basis, the local slope would be exactly $0$ for any $R$, as the antibonding $\phi_2(\rv)$ orbital which enters eq~\ref{eq:2elec2} has a node at the bond midpoint.

We also see in Figure~\ref{fig_slope_he} that the correlation energy density for the He isoelectronic series scales quickly towards an asymptotic constant as $Z$ increases. Furthermore, the local slope decays smoothly with distance from the nucleus, reflecting the behaviour of $v_{abij}(\mathbf{r})$. The behaviour for the Be isoelectronic series in Figure~\ref{fig_slope_be} is more complex. The KS HOMO-LUMO gap is known to increase\cite{SavColPol-IJQC-03} as $Z$ increases from $4$ to $10$, from which one would expect the correlation energy to become less negative according to the behaviour of $t_{ij}^{ab}$. In the core region this behaviour holds, however in the valence region the trend is opposite, with the correlation energy density becoming more negative with increasing $Z$. This suggests that the numerator of $t_{ij}^{ab}$ and the spatial dependence of $v_{abij}(\mathbf{r})$ due to the form of the KS orbitals are dominant in this region, provided that eq~\ref{eq:simplemp2} is sufficiently accurate for the Be isoelectronic series.
 
\subsection{The SCE model and the strong interaction limit}\label{sec_sce}
In recent years, the exact strong-coupling limit of the AC has been intensively studied.\cite{SeiGorSav-PRA-07,GorVigSei-JCTC-09,MalGor-PRL-12,MirSeiGor-JCTC-12,MirSeiGor-PRL-13,VucWagMirGor-JCTC} This limit reveals a new structure for the XC functional: instead of the traditional ingredients of DFAs (local density, density gradients, KS kinetic energy density, occupied and unoccupied KS orbitals) it is observed that certain {\em integrals of the density} appear in this limit, encoding highly non-local information.\cite{Sei-PRA-99,SeiGorSav-PRA-07,MalGor-PRL-12,MalMirCreReiGor-PRB-13,MenMalGor-PRB-14} 

Tests on model physical and chemical systems (electrons confined in low-dimensional geometries and low-density, ultracold dipolar systems, simple stretched bonds and anions) have shown\cite{MalGor-PRL-12,MalMirCreReiGor-PRB-13,MenMalGor-PRB-14,MalMirGieWagGor-PCCP-14,CheFriMen-JCTC-14,VucWagMirGor-JCTC,MalMirMenBjeKarReiGor-PRL-15} that taking into account this exact behaviour can pave the way for the solution of the strong correlation problem in DFT. However, the exact information encoded in the infinite coupling limit, described by the SCE functional, does not come for free: the SCE problem is ultra non-local, and, although sparse in principle, its non-linearity makes its exact evaluation for general three-dimensional geometry a complex task. A possible route to find suitable algorithms relies on the fact that constructing the exact SCE functional for a given density is equivalent to solving an optimal transport (or mass transportation theory) problem with a cost function given by the Coulomb interaction.\cite{ButDepGor-PRA-12,CotFriKlu-CPAM-13} This equivalence has triggered interest from the applied mathematics community working on optimal transport problems, which has led to the suggestion of several algorithms,\cite{MenLin-PRB-13,CheFriMen-JCTC-14,BenCarCutNenPey-SIAM-15,BenCarNen-arxiv-15} together with very interesting exact results.\cite{ColDepDim-CJM-15,DimGerNen-arxiv-15,ColStr-arxiv-15} 

So far, the SCE solution is known exactly for one-dimensional systems.\cite{ColDepDiM-CJM-14} For spherically symmetric systems, a conjectured solution\cite{SeiGorSav-PRA-07} that is very close to the exact one\cite{DiMGerNenSeiGor-xxx-15} (and it is in many cases, but not always,\cite{ColStr-arxiv-15} exact) has been proposed and used to address interesting physical problems.\cite{MenMalGor-PRB-13,MalMirMenBjeKarReiGor-PRL-15} Using algorithms and ideas from optimal transport, the SCE problem for the hydrogen molecule along the dissociation curve has just recently been solved and both the global\cite{CheFriMen-JCTC-14,VucWagMirGor-JCTC} and local\cite{VucWagMirGor-JCTC} SCE quantities have been computed. A more practical way to proceed is to build approximations for the SCE functional inspired by its exact form, as it was done in the construction of the already mentioned NLR functional.\cite{WagGor-PRA-14,ZhoBahErn-JCP-15} 

The SCE system complements the KS system.\cite{Sei-PRA-99,SeiGorSav-PRA-07,GorVigSei-JCTC-09} It corresponds to the wave function that minimizes the Hamiltonian of eq~\ref{eq:hamil_ac} when $\lambda \to \infty$. One can argue that the SCE system is a better starting point than the Kohn-Sham system for the description of very strongly correlated systems.\cite{MalMirCreReiGor-PRB-13,MenMalGor-PRB-14,CheFriMen-JCTC-14,VucWagMirGor-JCTC}

The SCE functional is defined as\cite{SeiGorSav-PRA-07,MalGor-PRL-12,MirSeiGor-JCTC-12}:
	\begin{align}
\mathcal{W}^{\rm SCE}[\rho]=\langle \Psi_{\infty}[\rho ]  | \hat{W}| \Psi_{\infty}[\rho ]\rangle.
		\label{eq:sce_func}
	\end{align}
The XC part $\mathcal{W}_{\text{xc},\infty}[\rho]$ can be easily extracted from $\mathcal{W}^{\rm SCE}[\rho]$, as $\mathcal{W}_{\text{xc},\infty}[\rho]=\mathcal{W}^{\rm SCE}[\rho]-U[\rho]$. The KS SCE approximation, proposed in ref~\citenum{MalGor-PRL-12}, uses the SCE functional to approximate the Hartree and exchange-correlation energy, and it is equivalent to setting $\mathcal{W}_\lambda[\rho]=\mathcal{W}_\infty[\rho]$ for all $\lambda$. It has been shown that KS SCE yields good energies for systems where correlation plays a dominant role, like electrons confined in low-density nanodevices or extremely stretched bonds. \cite{MenMalGor-PRB-14,MalMirCreReiGor-PRB-13,VucWagMirGor-JCTC,CheFriMen-JCTC-14,MalMirCreReiGor-PRB-13} On the other hand, KS SCE treats moderately and weakly correlated systems very poorly, giving energies that are unacceptably too low.\cite{MalMirGieWagGor-PCCP-14,VucWagMirGor-JCTC} A less drastic approximation is to construct a $\mathcal{W}_\lambda[\rho]$ model, in such a way that its $\lambda \to \infty$ limit is given by the exact or approximate value of $\mathcal{W}_\infty[\rho]$, as done in the pioneering work of Seidl {\em et. al.}\cite{SeiPerLev-PRA-99,SeiPerKur-PRL-00}  Analogously, one can also model $w_\lambda(\rv)$, imposing that its $\lambda \to \infty$ limit is given by $w_\infty(\rv)$. This latter approach is the main object of the following sections. 

In the SCE limit, the electrons are infinitely or perfectly correlated and their positions are given by an infinite superposition of classical configurations. The basic idea is that the electronic positions are all determined by a collective variable $\rv$, a feature that is captured by the so-called {\em co-motion functions} $\fv_i(\rv)$.\cite{SeiGorSav-PRA-07} If a reference electron is at $\rv$, then the position of all the other electrons in the system will be given by $\rv_i=\fv_{i}(\rv)$.\cite{SeiGorSav-PRA-07} Since the electrons are perfectly correlated, the probability of finding the reference electron at $\rv$ has to be the same as the probability of finding the $i$th electron at $\fv_{i}(\rv)$. Therefore, the co-motion functions have to satisfy the following differential equation:\cite{SeiGorSav-PRA-07}
	\begin{align}
		\rho(\fv_{i}(\rv))d\fv_{i}(\rv)=\rho(\rv)d\rv.
		\label{eq:dif_eq}
	\end{align}
For more details on the co-motion functions, including their group properties, see Refs.~\citenum{MirSeiGor-JCTC-12,ButDepGor-PRA-12,MenMalGor-PRB-13} and~\citenum{SeiGorSav-PRA-07}. 

In terms of the co-motion functions, the SCE functional $\mathcal{W}^{\rm SCE}[\rho]$ is given by\cite{MirSeiGor-JCTC-12}
\begin{align}
	\mathcal{W}^{\rm SCE}[\rho ]=\frac{1}{2}\int d\rv\rho(\rv)\sum_{i=2}^{N}\frac{1}{\left | \rv-\fv_i(\rv) \right|}.
\label{eq:vee_comotion}
	\end{align}
Despite the high nonlocality of the SCE functional, evident from eq~\ref{eq:dif_eq}, we can easily compute its functional derivative from the following expression\cite{MalGor-PRL-12,ButDepGor-PRA-12}
	\begin{align}
		\nabla v_{\rm SCE}(\rv)=-\sum_{i=2}^{N}\frac{\rv-\fv_i(\rv)}{\left | \rv-\fv_i(\rv) \right|^3}.
		\label{eq:pot_sce}
	\end{align}
 
Equation~\ref{eq:vee_comotion} suggests the following energy density in the SCE limit:
	\begin{align}
w_\infty(\rv)=\frac{1}{2}\sum_{i=2}^N\frac{1}{|\rv-\fv_i(\rv)|}-\frac{1}{2}v_{\text H}(\rv),
		\label{eq:sce_edens}
	\end{align}
where $v_{\text H}(\rv)$ is the Hartree potential. This expression is indeed in the gauge of the XC hole potential of eq~\ref{eq:wxc_def}, as proven in ref~\citenum{MirSeiGor-JCTC-12}. Being derived from a wavefunction, the $w_\infty(\rv)$ energy density decays like $\sim -\frac{1}{2 |\rv|}$, similar to the physical  ($\lambda=1$) and the exchange ($\lambda=0$) energy densities of eq~\ref{eq:lieb_wxc}. Its functional derivative, eq~\eqref{eq:pot_sce}, has also the correct asymptotic $v_{\text{xc}}\sim -\frac{1}{|\rv|}$ behaviour. 

To solve the SCE problem for spherically symmetric systems (the He and Be isoelectronic series considered in this paper) we have used the approach presented in ref~\citenum{SeiGorSav-PRA-07}, which is exact if $N=2$. For atomic densities with $N>2$ it could be either a very good approximation for the true minimum of eq~\ref{eq:sce_func}, or again, the exact result.\cite{ColStr-arxiv-15,DiMGerNenSeiGor-xxx-15} For the H$_2$ molecule we have used the results of ref~\citenum{VucWagMirGor-JCTC}, where the SCE energy density has been computed by obtaining the co-motion function from the dual Kantorovich formulation\cite{ButDepGor-PRA-12,Dep-arxiv-15} of the SCE problem.

\subsection{Local interpolation models}\label{sec_locforms}
The local interpolation models tested in this work are largely simple translations of the well--established global interpolation models into a local form. This was done for the model of Seidl, Perdew and Levy (SPL),\cite{SeiPerLev-PRA-99} the ``simplified'' model of Liu and Burke,\cite{LiuBur-PRA-09} which will be referred here as the LB model and the Pad{\'e}$[1/1]$ model.\cite{Ern-CPL-96,SanCohYan-JCP-06} Each of the energy densities resulting from the three mentioned models is constructed from three local parameters, $a$, $b$ and $c$, which are defined in the gauge of the XC--hole. The functional forms of these three models are summarized in Table~\ref{tab_ac_eqns}. 

In addition to these, we constructed a local form of the two--legged representation\cite{BurErnPer-CPL-97} which, given some value of $w_{1}(\mathbf{r})$, takes the form
\begin{subequations}\begin{align}
w_{\lambda}(\mathbf{r}) &= 
\begin{cases}
w_{0}(\mathbf{r}) + \lambda w'_{0}(\mathbf{r}), & \lambda \leqslant x_{\lambda} \\
w_{1}(\mathbf{r}), & \lambda > x_{\lambda}
\end{cases} \label{eq:2leg} \\[2ex]
x_{\lambda} &= \frac{w_{1}(\mathbf{r}) - w_{0}(\mathbf{r})}{w'_{0}(\mathbf{r})}. \label{eq:xlam}
\end{align}\end{subequations}

Whenever we used the two--legged representation to model the local AC in this work, we did it by incorporating the interpolated $w_{1}(\mathbf{r})$ of the LB model: $w_{1}(\rv) \approx w_{1}^{\rm LB}(\rv)$. By doing the local interpolation this way, we use the following three input quantities: $w_0(\rv)$, $w_0'(\rv)$ and $w_\infty(\rv)$ and circumvent the direct utilization of the full interacting energy density, $w_{1}(\mathbf{r})$.
In each of these four models, integration of $w_{\lambda}(\mathbf{r})$ with respect to coupling--constant gives the $\lambda$--averaged energy density $\bar{w}_{\rm xc}(\mathbf{r})$ which, if spatially integrated according to eq~\ref{eq:ac_wxc}, yields the XC--energy $E_{\rm xc}[\rho]$.

An important observation in the translation of global to local models is that, whilst the following global inequalities are always satisfied,
\begin{equation}\label{eq:ineqglo}
\mathcal{W}_{0}[\rho] \geq E_{\rm xc}[\rho] \geq \mathcal{W}_{1}[\rho] \geq \mathcal{W}_{\infty}[\rho],
\end{equation}
their local counterparts do not necessarily satisfy these same inequalities. It has previously been observed for the Hooke's atom series that, in the tail regions of the density, $w_{\infty}(\mathbf{r})$ can be less negative than $w_{1}(\mathbf{r})$.\cite{MirSeiGor-JCTC-12} In this work, the crossing of $w_{\infty}(\mathbf{r})$ with $\bar{w}_{\rm xc}(\mathbf{r})$, $w_{1}(\mathbf{r})$ and $w_{0}(\mathbf{r})$ has only been observed in the tail regions of the density and is thought to be an artefact of the numerical instability that occurs where the density is very small. 

\begin{table*}
\caption{The mathematical forms of the local AC interpolation models (for the Pad{\'e}$[1/1]$ model, $p > 0$, $p \in \mathbb{R}$).}\label{tab_ac_eqns}
\centering
\begin{tabular*}{0.9\textwidth}{l@{\extracolsep{\fill}}mmmmc} \hline\hline\noalign{\vskip 1ex}
  & w_{\lambda}(\mathbf{r}) & a(\mathbf{r}) & b(\mathbf{r}) & c(\mathbf{r}) & Refs. \\[1ex]\hline\noalign{\vskip 1ex}
\textbf{SPL} & a + \frac{b}{\sqrt{1 + c \lambda}} & w_{\infty}(\mathbf{r}) & w_{0}(\mathbf{r}) - w_{\infty}(\mathbf{r}) & -\frac{2 w_{0}'(\mathbf{r})}{w_{0}(\mathbf{r}) - w_{\infty}(\mathbf{r})} & \citenum{SeiPerLev-PRA-99,SeiGorSav-PRA-07} \\[4ex]
\textbf{LB} & a + b \left(\frac{1}{(1 + c \lambda)^{2}} + \frac{1}{\sqrt{1 + c \lambda}}\right) & w_{\infty}(\mathbf{r}) & (w_{0}(\mathbf{r}) - w_{\infty}(\mathbf{r}))/2 & -\frac{4 w_{0}'(\mathbf{r})}{5(w_{0}(\mathbf{r}) - w_{\infty}(\mathbf{r}))} & \citenum{LiuBur-PRA-09} \\[4ex]
\textbf{Pad\'e}[1/1] & a + \frac{b \lambda}{1 + c \lambda} & w_{0}(\mathbf{r}) &  w_{0}'(\mathbf{r}) & \frac{-w_{0}(\mathbf{r}) + w_{p}(\mathbf{r}) - w_{0}'(\mathbf{r})}{w_{0}(\mathbf{r}) - w_{p}(\mathbf{r})} & \citenum{Ern-CPL-96,SanCohYan-JCP-06} \\[2ex]\hline\hline
\end{tabular*}
\end{table*}
\section{Results}\label{sec_results}
\subsection{Helium isoelectronic series}\label{sec_he_ser}
Although the helium isoelectronic series is a set of only two--electron systems, it is a useful series to consider in evaluating the local interpolation models as most standard DFAs incorrectly characterize the hydride ion (H$^-$), failing to predict its existence as a bound electronic system.\cite{MirUmrMorGor-JCP-14,PeaMilTeaToz-JCP-08} Here, local interpolation models are constructed from energy densities acquired by the Lieb maximisation at the FCI level, as described in section~\ref{liebmax}, in the range $0 \leq \lambda \leq 1$ and at $\lambda = \infty$ by evaluating the SCE functional on the $\lambda=1$ density, also at the FCI level of theory. 
\begin{table}
\caption{Reference and interpolated $E_{c}$ values, in Hartree, for the He isoelectronic series.}\label{tab_he}
\centering
\begin{tabular}{@{}lrrrrrr@{}} \hline\hline\noalign{\vskip 0.25ex}
Z  &   FCI   & local SPL & global SPL & local LB &  Pad{\'e}$[1/1]$ & local 2--leg \\ \hline
1  & -0.0409 & -0.0367   & -0.0368    & -0.0398  & -0.0401 & -0.0477               \\
2  & -0.0400 & -0.0378   & -0.0380    & -0.0394  & -0.0399    & -0.0435           \\
3  & -0.0410 & -0.0393   & -0.0395    & -0.0404  & -0.0409 & -0.0431               \\
4  & -0.0416 & -0.0402   & -0.0404    & -0.0411  & -0.0415 & -0.0433              \\
5  & -0.0418 & -0.0408   & -0.0409    & -0.0415  & -0.0418 & -0.0433              \\
6  & -0.0419 & -0.0410   & -0.0411    & -0.0416  & -0.0418 & -0.0431               \\
7  & -0.0414 & -0.0407   & -0.0408    & -0.0412  & -0.0414 & -0.0423              \\
8  & -0.0412 & -0.0405   & -0.0406    & -0.0410  & -0.0412 & -0.0420             \\
9  & -0.0411 & -0.0405   & -0.0406    & -0.0409  & -0.0411 & -0.0419              \\
10 & -0.0411 & -0.0405   & -0.0407    & -0.0408  & -0.0411 & -0.0418               \\ \hline\hline
\end{tabular}
\end{table}

In Table~\ref{tab_he}, the correlation energies given by local forms of the SPL, LB, two-legged representation (the column labelled ``2--leg'') and Pad\'e$[1/1]$ models (the latter parameterized using the accurate values for $w_{1}(\mathbf{r})$, in order to compare with models that, instead, use the $\lambda\to\infty$ information) are given, along with that given by the global SPL model and the FCI correlation energy for comparison. This data shows that the local interpolation correlation energies are in close agreement with the FCI reference values; the mean absolute errors (MAE) of the local interpolation models are $2.0$~mH, $1.5$~mH, $0.5$~mH and $0.1$~mH, for the two-legged representation, SPL, LB and Pad{\'e}$[1/1]$ models, respectively.

As would be expected, the local Pad{\'e}$[1/1]$ is the most accurate of the models, given that it is derived from the full interacting energy density. This data further suggests that the local LB model is marginally superior to the local SPL and the two-legged representation. However, comparing the global and local models shows a slightly lower error for the global model; the local SPL model has an MAE of $1.5$~mH, compared to $1.3$~mH for the global model.
\begin{figure}
\includegraphics[width=\linewidth]{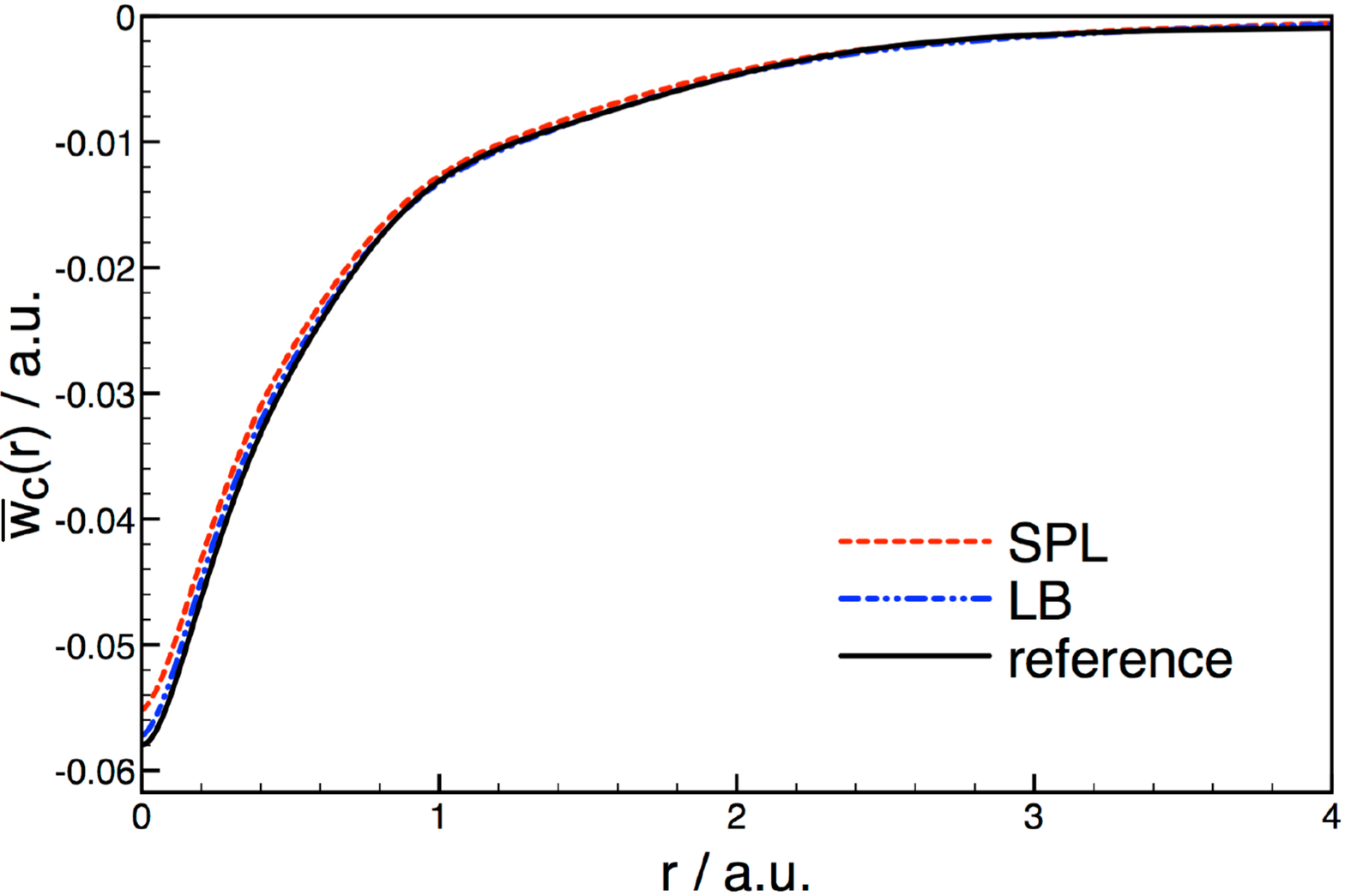}
\caption{Plots comparing the FCI, local LB and local SPL $\lambda$--averaged correlation energy density in the helium atom.}
\label{fig_w_he}
\end{figure}

Figure~\ref{fig_w_he} compares the FCI $\bar{w}_{\rm c}(\mathbf{r})$ with that of the local LB and SPL models, for the helium atom. This reflects the numerical data in Table~\ref{tab_he}, both being very close to the FCI energy density but with slightly lower error in the LB model. 
 
\subsection{Beryllium isoelectronic series}\label{sec_be}
The changes in correlation energy across the beryllium isoelectronic series are somewhat more complicated than those in the helium isoelectronic series, and its explanation involves the interplay of several effects. With increasing nuclear charge, the density becomes increasingly contracted, suggesting that the correlation energy should approach the high--density limit for very large $Z$. However, this is accompanied by a changing KS HOMO--LUMO gap, here the energy difference between $2s$ and $2p$ orbitals, which increases from $Z=4\to13$ before decreasing with higher $Z$ values.\cite{SavColPol-IJQC-03}
\begin{table}
\centering
\caption{Reference and interpolated $E_{c}$ values, in Hartree, for the Be isoelectronic series.}\label{tab_be}
\begin{tabular}{@{}lrrrrrr@{}} \hline\hline\noalign{\vskip 0.25ex}
Z  &   CCSD  & local SPL & global SPL & local LB & Pad{\'e}$[1/1]$ & local 2--leg \\ \hline
4  & -0.0920 & -0.0876   & -0.1049    & -0.0925  & -0.0911 & -0.1046              \\
5  & -0.1089 & -0.1041   & -0.1250    & -0.1100  & -0.1076 & -0.1246               \\
6  & -0.1244 & -0.1202   & -0.1455    & -0.1271  & -0.1229 & -0.1444           \\
7  & -0.1389 & -0.1363   & -0.1668    & -0.1443  & -0.1373 & -0.1645               \\
8  & -0.1534 & -0.1532   & -0.1898    & -0.1626  & -0.1517 & -0.1859               \\
9  & -0.1683 & -0.1717   & -0.2157    & -0.1826  & -0.1666 & -0.2098             \\
10 & -0.1833 & -0.1920   & -0.2447    & -0.2046  & -0.1817 & -0.2361              \\ \hline\hline
\end{tabular}
\end{table}

Table~\ref{tab_be} shows the reference and interpolated $E_{c}$ results for the Be isoelectronic series, with $Z$ in the range $4-10$. The wave function for $\lambda$ values between $0$ and $1$ has been computed in the same way as for the He isoelectronic series, however at the CCSD level of theory rather than FCI. As for the helium series, the local Pad{\'e}$[1/1]$ that uses $w_{1}(\mathbf{r})$ is the most accurate of the local interpolation models. However, in contrast to the findings for He isoelectronic series, the local interpolation models are much more accurate than the global models. For example, in the case of F$^{5+}$ the global SPL model has an MAE of $47.4$~mH, whereas the error for the local SPL model is $3.5$~mH. The local two-legged representation interpolation underestimates the correlation energies of the elements of the given series. We discuss in more details this model in the next subsection.
\begin{figure}
\includegraphics[width=\linewidth]{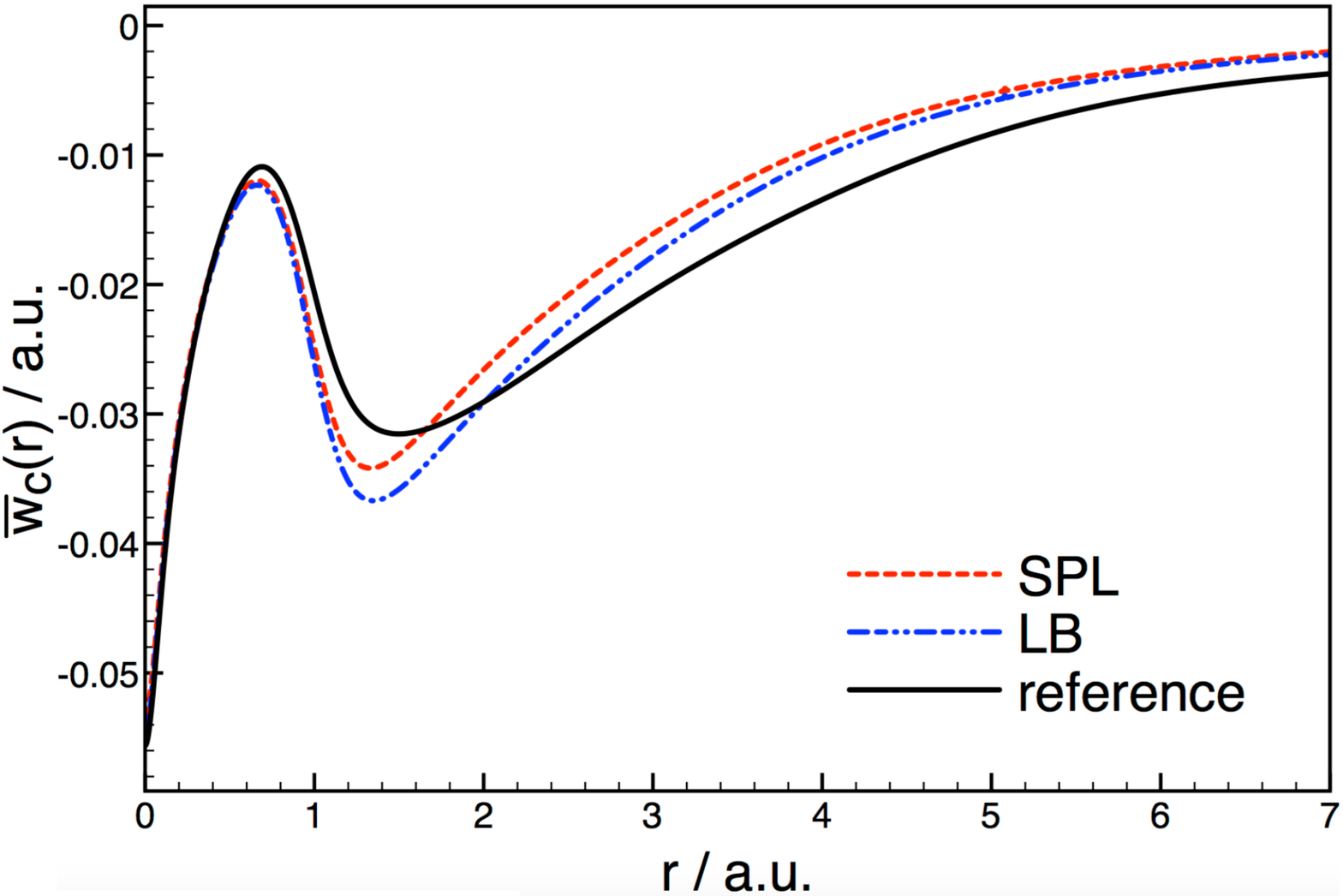}
\caption{Plots comparing the CCSD, local LB and local SPL $\lambda$--averaged correlation energy density in the beryllium atom.}
\label{fig_w_be}
\end{figure}

Figure~\ref{fig_w_be} shows the $\lambda$--averaged correlation energy densities for the beryllium atom. The shape of $\bar{w}_{\rm c}(\mathbf{r})$ reflects the shell structure of the Be atom.\cite{ColSav-JCP-99,IroTea-MP-15} The local SPL and LB interpolation models appear to qualitatively capture the shell structure of $\bar{w}_{\rm c}(\mathbf{r})$, however in some regions it overestimates the reference value whilst in other regions the converse is the case. The error cancellation that results from this is the most likely explanation for the superior accuracy of the local models in comparison to the global models. 

\subsection{Hydrogen molecule}\label{sec_h2}
Despite the development of DFT into the most widely--applied electronic structure method, and the wealth of XC--DFAs that have been developed, there are some systems for which no combination of DFAs provide an accurate description. A well--known example of such a system is the dissociating $\rm H_{2}$ molecule.\cite{FucNiqGonBur-JCP-05,VucWagMirGor-JCTC-15} Standard DFAs become increasingly inaccurate with greater H--H bond length, reflecting a fundamental flaw of DFAs in their inability to properly treat strong correlation.  

\begin{figure*}
\includegraphics[width=\textwidth]{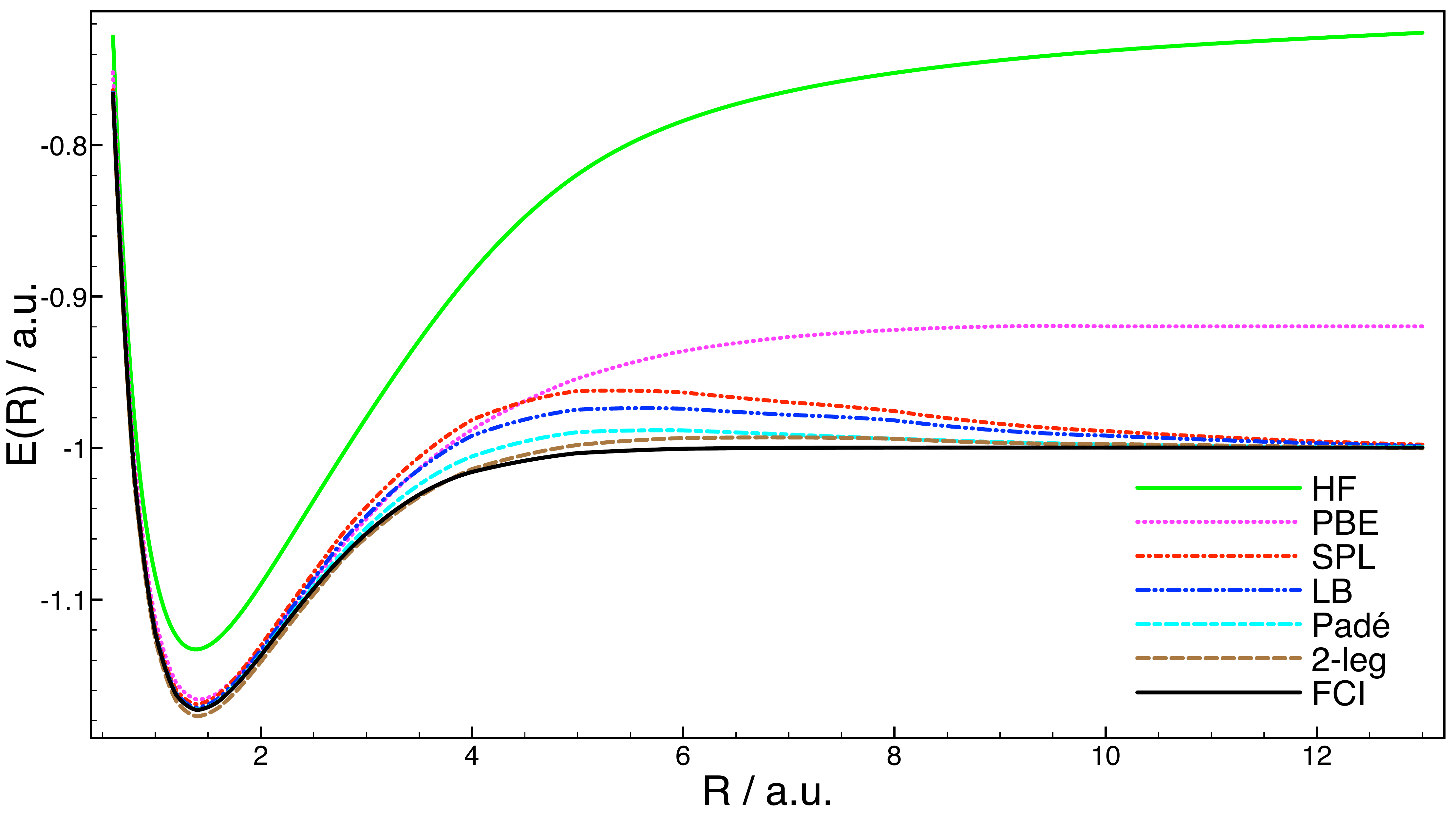}
\caption{Potential energy curves for the H$_2$ molecule with internuclear distance $R$ / a.u., which are obtained using the local interpolation methods: SPL, Liu-Burke, two-legged representation combined with the Liu-Burke model, Pad{\'e}$[1/1]$ with $w_{1}(\mathbf{r})$. Restricted HF, PBE and FCI curves are also shown for comparison.}
\label{fig_h2curves}
\end{figure*}

It has been seen previously\cite{CheFriMen-JCTC-14,VucWagMirGor-JCTC} that KS SCE correctly predicts the dissociation of $\rm H_{2}$ in a spin--restricted formalism, however at equilibrium geometry the energies it predicts are extremely low and the bond lengths predicted are overly short. The overall accuracy of KS-SCE for $\rm H_{2}$ dissociation can be substantially improved by the addition of nonlocal corrections.\cite{VucWagMirGor-JCTC}

Figure~\ref{fig_h2curves} shows the dissociation curves for $\rm H_{2}$ given by the local interpolation models, along with those given by HF, FCI and the PBE functional\cite{PerBurErn-PRL-96} for comparison. The computational details are the same as those of the He isoelectronic series, and the PBE, HF and FCI curves have been obtained from the \textsc{Dalton} quantum chemistry package\cite{Dalton-WIRES-14} all within the uncontracted aug-cc-pCVTZ basis set.\cite{Dun-JCP-89} The SCE energy density has been computed by using the dual Kantorovich method.\cite{VucWagMirGor-JCTC} 

It can be seen in Figure~\ref{fig_h2curves} that all of the interpolation models correctly predict the dissociation of $\rm H_{2}$, which follows from their inclusion of $w_{\infty}(\rv)$. In global AC models, at infinite separation the initial slope diverges as a result of the vanishing HOMO--LUMO gap, and the SPL and LB models reduce to $\mathcal{W}_{\infty}[\rho]$, yielding the exact energies. However, the dissociation curves produced by the local models approach the FCI curve slowly, resulting in an unphysical `bump'--like feature. This is a well--known failing of DFT, having been observed with other functionals, such as the random--phase approximation \cite{FucNiqGonBur-JCP-05} and even the global Pad{\'e}$[1/1]$ model with $\mathcal{W}_{1}[\rho]$.\cite{PeaTeaToz-JCP-07} It can be seen in Figure~\ref{fig_h2curves} that this is not remedied by the local interpolation approach, as the curve obtained by the local Pad{\'e}$[1/1]$ also exhibits this unphysical bump, as does that given by the local SPL model and, to a lesser extent, the local LB model.
\begin{figure}
\includegraphics[width=9cm]{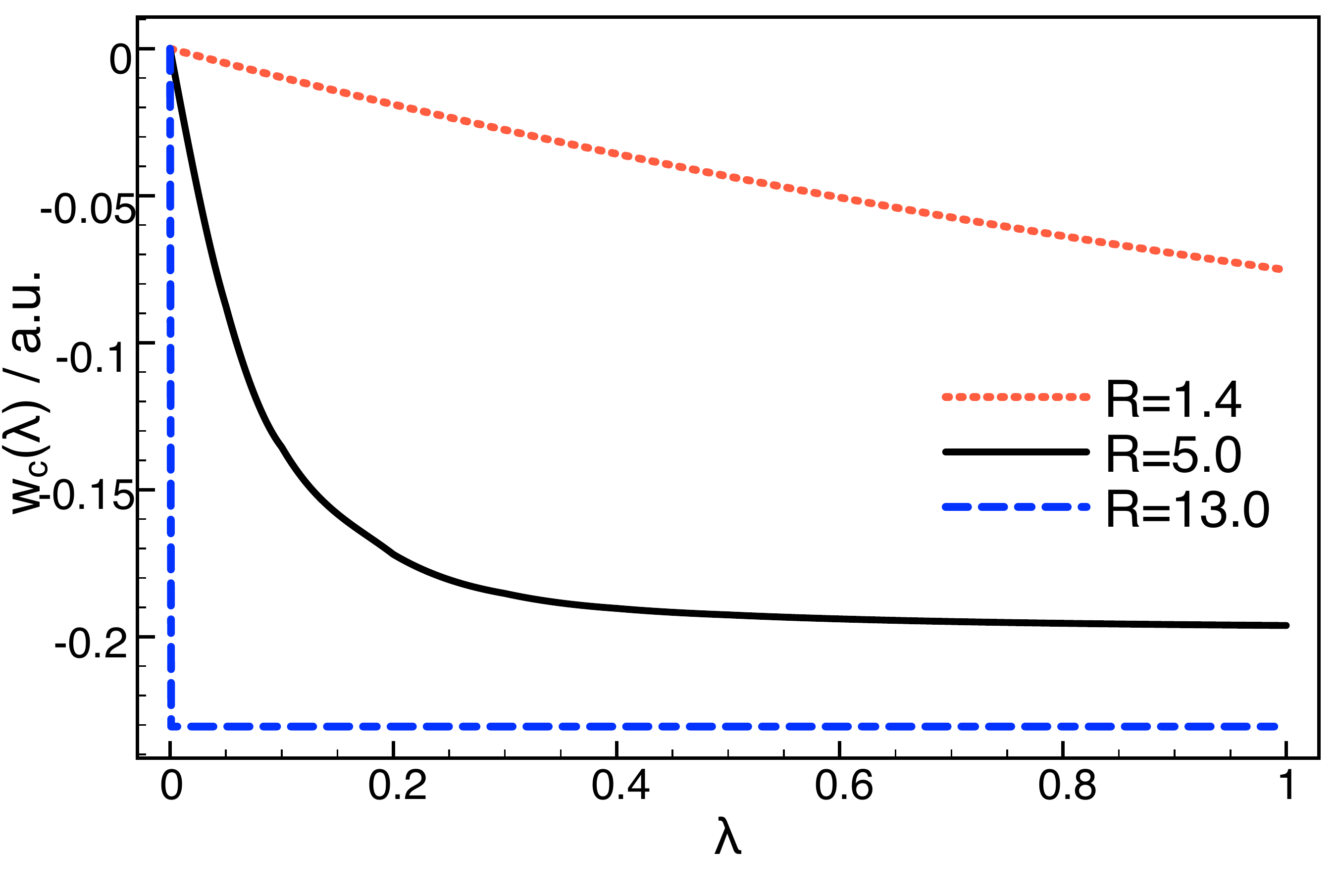}
\caption{The FCI local correlation AC curves at one of the nuclei of H$_2$ for different for different internuclear separations, $R$.}
\label{fig_h2_acs}
\end{figure}
\begin{figure}
\includegraphics[width=9cm]{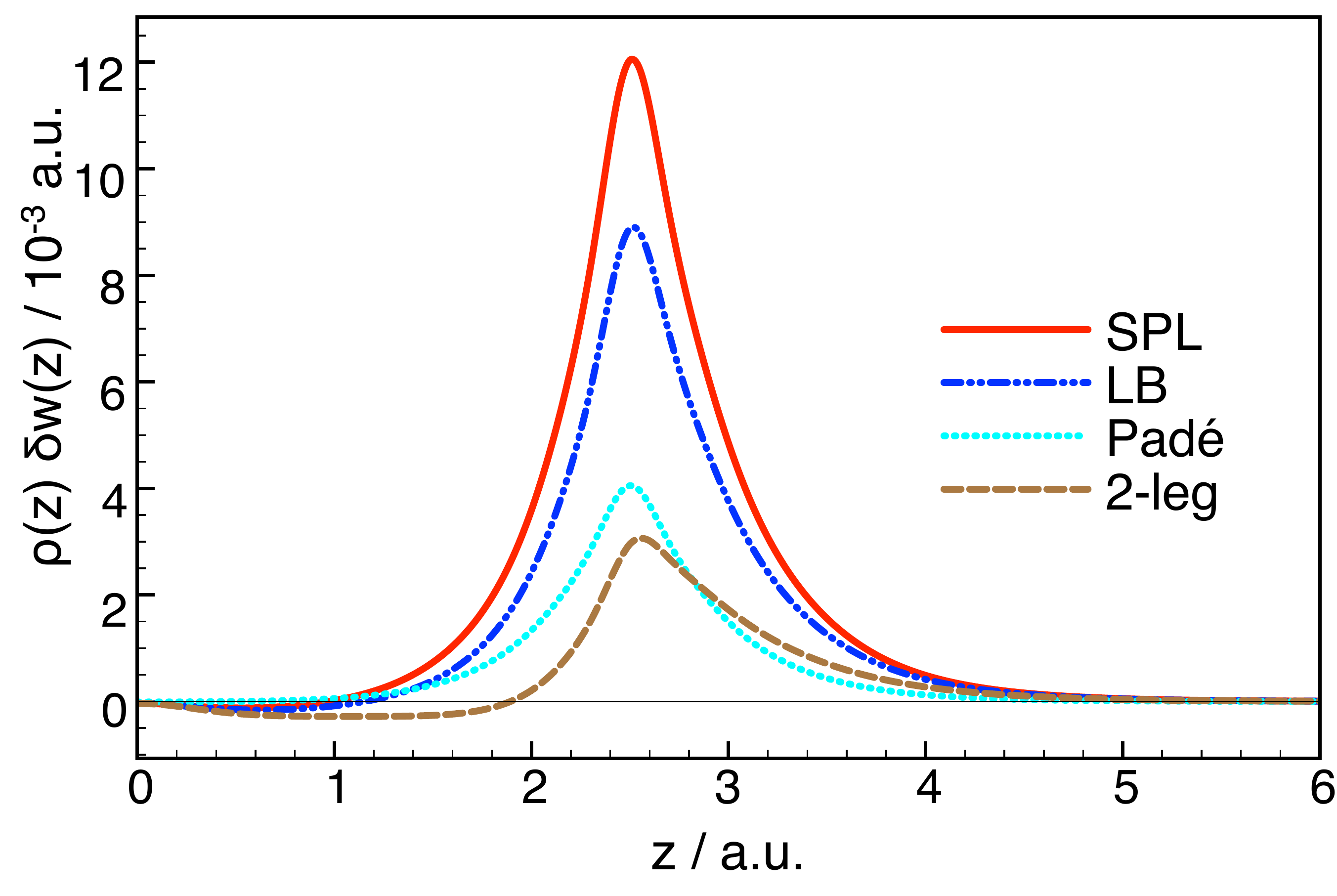}
\includegraphics[width=9cm]{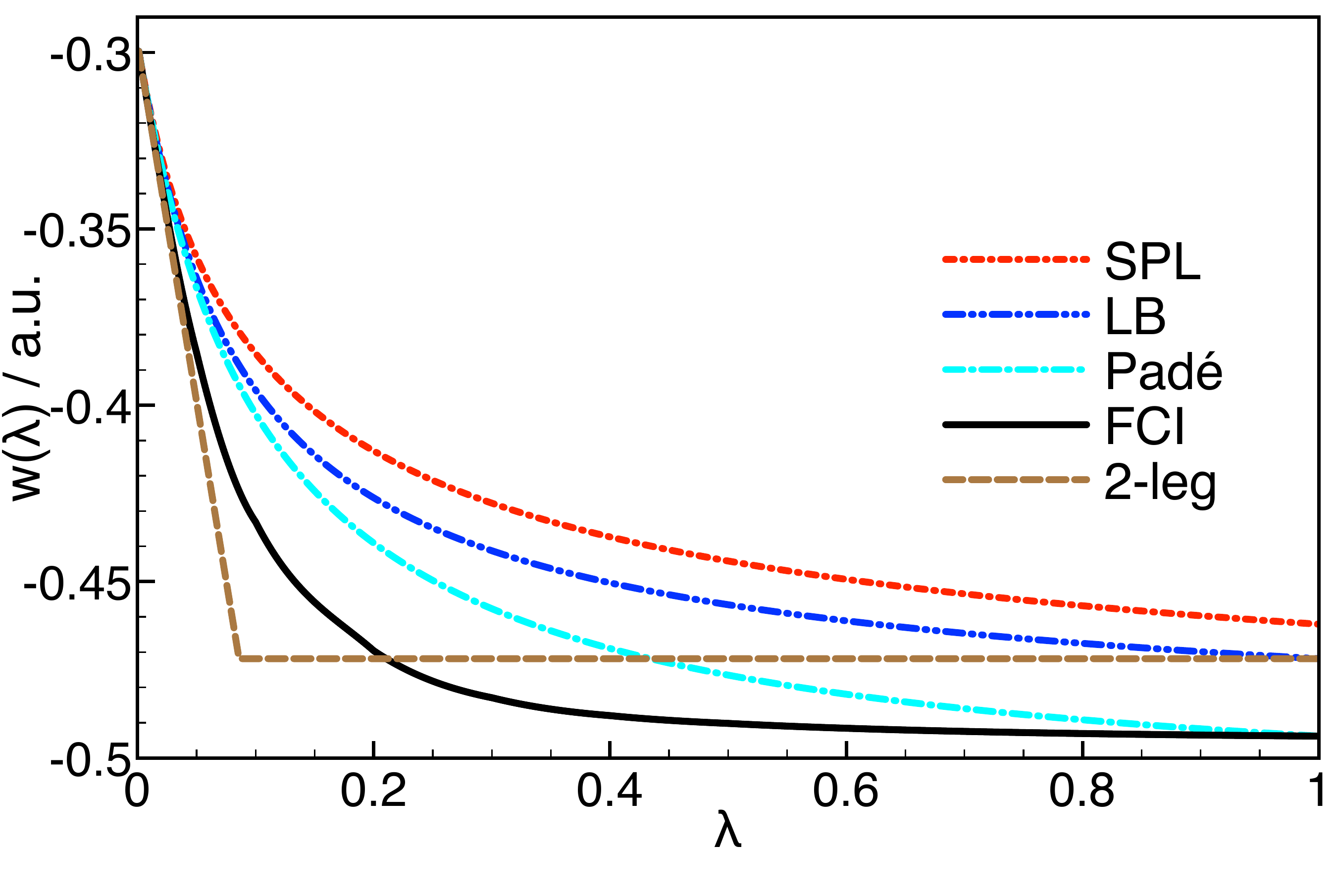}
\caption{Plots of the difference between FCI and interpolated $\lambda$--averaged energy densities, $\delta w(z) = \bar{w}_{\rm FCI}(z)- \bar{w}_{\rm model}(z)$, with respect to the distance from the bond midpoint, $z$ / a.u. (upper panel), and the local AC curves at one of the nuclei of the FCI and local interpolation models (lower panel), both in H$_2$ with a $5.0$ a.u. bond length.}
\label{fig_h2_5}
\end{figure}

To analyse why the intermediate region is less accurately described by the local interpolation methods than the equilibrium and stretched region, we show in Figure~\ref{fig_h2_5} the correlation component of the local AC at one of the nuclei of the hydrogen molecule at different bond lengths: $R=1.4$ a.u. (at equilibrium), $R=5.0$ a.u. (the intermediate region) and $R=13.0$ a.u. (stretched bond). The structure of the three local AC curves at one of the nuclei is very similar to the structure of the corresponding global AC curves.\cite{TeaCorHel-JCP-09} From the given figure we see that at equilibrium the local AC is almost linear, so we can expect that even a single line segment approximation to the local AC: $w_\lambda(\rv) \approx w_0(\rv)+w'_0(\rv)$ would properly capture the shape of the given local AC curve. The local AC curve at the nuclei of the stretched H$_2$ exhibits the characteristic `L-shape', which was also observed in the case of the corresponding global AC curve.\cite{TeaCorHel-JCP-09} We would expect that the two-legged representation would capture the given local AC very well, but even a single line segment approximation: $w_\lambda(\rv) \approx w_\infty(\rv)$, this time coming from the strong coupling limit, would be highly accurate for the stretched H$_2$.\cite{VucWagMirGor-JCTC} In contrast to the local AC curves of the stretched and H$_2$ at equilibrium, the curvature of the local AC curve at the intermediate bond length is highly pronounced. The shapes of the local AC curves at the nuclei mirror the difference in correlation regimes present in the hydrogen molecule at different bond lengths. While in the H$_2$ at equilibrium and at very stretched bond length, correlation is almost purely dynamical and almost purely static, respectively, in the intermediate dissociation region there is a subtle interplay between the dynamical and static correlation.

In the intermediate region of the dissociation curve, where the unphysical bump is present, the local two-legged representation model is more accurate than the local Pad{\'e}$[1/1]$ which we always use here with $w_{1}(\mathbf{r})$. This may be understood by comparing the exact local AC data with the interpolated quantities. The top panel of Figure~\ref{fig_h2_5} shows the difference between $\bar{w}_{\rm FCI}(\mathbf{r})$ and that of each of the local interpolation models, along the H--H bond at the $5.0$ a.u. geometry, as a function of the distance from the bond midpoint $z$. This difference $\delta w(\mathbf{r}) = \bar{w}_{\rm FCI}(\mathbf{r}) - \bar{w}_{\rm model}(\mathbf{r})$, is multiplied by the density to represent an energy per volume element. It shows that the local SPL energy density is the one that most overestimates the $\bar{w}(\mathbf{r})$. The error is smaller for the LB model and even more so for the local Pad{\'e}$[1/1]$ model. The error is smallest in the two--legged model, obtained using the $w_{1}(\mathbf{r})$ of the local LB. Furthermore, there is the error cancellation in the two--legged model, as there are regions where the $\bar{w}(\mathbf{r})$ of this model underestimates $\bar{w}_{\rm FCI}(\mathbf{r})$.

It can also be seen that the curves shown in the top panel of Figure~\ref{fig_h2_5} have a maximum at the nucleus ($z=2.5$). Focusing on this region, it appears that the FCI curve meets that of the Pad{\'e}$[1/1]$ at $\lambda=1$, and that the two--legged representation curve meets that of the LB model also at $\lambda=1$. This follows from the construction of the Pad{\'e}$[1/1]$ and two--legged curves from $w_{1}^{\rm FCI}(\mathbf{r})$ and $w_{1}^{\rm LB}(\mathbf{r})$ respectively. All curves, except for that of the two-legged model, lie above the FCI curve. In the case of the two--legged interpolation model, the first line segment is below the FCI curve, as a result of eq~\ref{eq:2leg} and the convexity of the given local AC curve. The second line segment that starts at $x_{\lambda} \sim 0.1$ is given by $w_{1}^{\rm LB}(\mathbf{r})$, and lies above the FCI curve. The resulting error cancellation makes it clear why the two--legged representation appears more accurate than the other models.

\section{Conclusion and Perspectives}\label{sec_concl}
In this work we have studied local interpolations along the adiabatic connection for the He and Be isoelectronic series and the hydrogen molecule, by using accurate input local quantities computed in the gauge of the electrostatic potential of the XC hole, and comparing the results with nearly exact energy densities defined in the same way. In order to obtain approximations to the local AC over the physical regime ($0 \leq \lambda \leq 1$), we constructed interpolation models between the weak and strong coupling limits of DFT. The weak coupling energy densities were obtained using the Lieb variation principle, whilst the strong coupling limit energy densities were obtained using the strictly-correlated electrons (SCE) approach. The inclusion of the SCE information in density functional approximations helps to ensure their ability to capture the strong correlation effects. 

Unlike previous attempts in this direction that used global (integrated over all space) input quantities to model the AC, the local approach is more amenable to the construction of approximations that do not violate size consistency, at least in the usual DFT sense.\cite{GorSav-JPCS-08,Sav-CP-09} Since the aim here is to work in a restricted formalism, avoiding to mimic strong correlation with symmetry breaking, some care must be taken when discussing size consistency. In fact, strictly speaking, in a restricted framework the energy densities of the second-order perturbation theory and exact exchange are not intensive quantities in the presence of near degeneracy,\cite{GorSav-JPCS-08,Sav-CP-09} which is the main challenge of capturing strong correlation within DFT.\cite{Sav-INC-96,CohMorYan-SCI-08,MorCohYan-PRL-09}

In future work we will test different approximations for the SCE energy densities and the local slope. The development of algorithms for solution of the SCE problem is a very active research field. In spite of the recent improvements, we still lack an algorithm that will solve the SCE problem for general 3D molecular geometries at low computational cost. However, a good candidate to approximate the SCE energy density in the gauge of the XC hole potential is the nonlocal radius functional (NLR),\cite{WagGor-PRA-14} which has been already implemented and used in ref~\citenum{ZhoBahErn-JCP-15}.

In addition to numerically exploring the local AC we have also reported the local weak-coupling slope of the adiabatic connection and derived an approximate expression for it in terms of occupied and unoccupied orbitals. This quantity is very important to signal the amount of correlation at each point of space. In our future work we will implement this expression and test it against the results reported here.

\section*{Acknowledgements}\label{ack}
We are very grateful to Derk Kooi and Andr{\'e} Mirtschink for a critical reading of the manuscript and insightful suggestions to improve it. We acknowledge financial support from the European Research Council under H2020/ERC Consolidator Grant corr-DFT (Grant No. 648932) and the Netherlands Organization for Scientific Research (NWO) through an ECHO grant (717.013.004). A. M. T. is grateful for support from the Royal Society University Research Fellowship scheme. A. M. T and T. J. P. I. are grateful for support from the Engineering and Physical Sciences Research Council (EPSRC), (Grant No. EP/M029131/1). We are grateful for access to the University of Nottingham High Performance Computing Facility.

\bibliography{biblioPaola,biblio_spec,biblio1,biblio_add}

\end{document}